# Energy surface, chemical potentials, Kohn-Sham energies in spin-polarized density functional theory

T. Gál* and P. Geerlings

Department of General Chemistry (ALGC), Member of the QCMM Alliance Ghent-Brussels,
Free University of Brussels (VUB), Pleinlaan 2, 1050 Brussel, Belgium

**Abstract:** On the basis of the zero-temperature grand canonical ensemble generalization of the energy $E[N,N_s,v,B]$ for fractional particle $N$ and spin $N_s$ numbers, the energy surface over the $(N,N_s)$ plane is displayed and analyzed in the case of homogeneous external magnetic fields $B(\vec{r})$. The (negative of the) left/right-side derivatives of the energy with respect to $N$, $N_\uparrow$, and $N_\downarrow$ give the fixed-$N_s$, spin-up, and spin-down ionization potentials/electron affinities, respectively, while the derivative of $E[N,N_s,v,B]$ with respect to $N_s$ gives the (signed) half excitation energy to a state with $N_s$ increased (or decreased) by 2. The highest occupied and lowest unoccupied Kohn-Sham spin-orbital energies are identified as the corresponding spin-up and spin-down ionization potentials and electron affinities. The excitation energies to the states with $N_s \pm 2$ can be obtained as the differences between the lowest unoccupied and the opposite-spin highest occupied spin-orbital energies, if the $(N,N_s)$ representation of the Kohn-Sham spin-potentials is used. The cases where the convexity condition on the energy does not hold are also discussed. Finally, the discontinuities of the energy derivatives and the Kohn-Sham potential are analyzed and related.

*Email address: galt@phys.unideb.hu

# I. Introduction

The great success of density functional theory (DFT) [1,2] in quantum chemistry and solid-state physics stems from its use of the electron density as basic variable in the place of the complicated many-variable, complex wavefunction. The cornerstone of DFT is the fact that there exists a functional

$$E_v[n] = F[n] + \int n(\vec{r}) v(\vec{r}) d\vec{r} \qquad (1)$$

of the electron density $n(\vec{r})$ whose minimum over $n(\vec{r})$'s of a given norm $N$,

$$N = \int n(\vec{r}) d\vec{r} , \qquad (2)$$

delivers the ground-state energy of an $N$-electron system in a given external potential $v(\vec{r})$,

$$E[N,v] = \min_{n \mapsto N} \left\{ F[n] + \int n(\vec{r}) v(\vec{r}) d\vec{r} \right\} , \qquad (3)$$

and the minimizing $n(\vec{r})$ is the ground-state density of the system [3-5]. This minimization principle leads to the Euler-Lagrange equation

$$\frac{\delta F[n]}{\delta n(\vec{r})} + v(\vec{r}) = \mu \qquad (4)$$

for the ground-state density in $v(\vec{r})$, with the Lagrange multiplier $\mu$ corresponding to the conservation constraint on the norm Eq.(2) of $n(\vec{r})$.

A spin-resolved version of DFT has also been developed [6,7] (see also [8-10]), with the spin-up $n_\uparrow(\vec{r})$ and spin-down $n_\downarrow(\vec{r})$ densities, or the total electron density $n(\vec{r})$ plus the spin-polarization density $n_s(\vec{r})$, as basic variables. The increased degree of freedom due to the new variable enables spin-polarized DFT to treat an additional, (collinear) magnetic external field $B(\vec{r})$ beside the electrostatic potential $v(\vec{r})$, and as recognized later (see, e.g., [9,11]), the lowest-lying state with any given spin number

$$N_s = \int n_s(\vec{r}) d\vec{r} , \qquad (5)$$

i.e., not only the ground state. The corresponding generalization of the minimization principle Eq.(3) is

$$E[N, N_s, v, B] = \min_{n \mapsto N, n_s \mapsto N_s} \left\{ F[n, n_s] + \int n(\vec{r}) v(\vec{r}) d\vec{r} - \int n_s(\vec{r}) \beta_e B(\vec{r}) d\vec{r} \right\} . \qquad (6)$$

The Euler-Lagrange equations arising from Eq.(6) for the determination of the lowest-lying state with given $N$ and $N_s$ in external fields $v(\vec{r})$ and $B(\vec{r})$ are

$$\frac{\delta F[n,n_s]}{\delta n(\vec{r})} + v(\vec{r}) = \mu \tag{7a}$$

and

$$\frac{\delta F[n,n_s]}{\delta n_s(\vec{r})} - \beta_e B(\vec{r}) = \mu_s, \tag{7b}$$

in the $(N, N_s)$ representation, or

$$\frac{\delta F[n_\uparrow, n_\downarrow]}{\delta n_\uparrow(\vec{r})} + v(\vec{r}) - \beta_e B(\vec{r}) = \mu_\uparrow, \tag{8a}$$

and

$$\frac{\delta F[n_\uparrow, n_\downarrow]}{\delta n_\downarrow(\vec{r})} + v(\vec{r}) + \beta_e B(\vec{r}) = \mu_\downarrow, \tag{8b}$$

in the $(N_\uparrow, N_\downarrow)$ representation. The Lagrange multipliers $\mu$, $\mu_s$, or $\mu_\uparrow$, $\mu_\downarrow$, emerge from the fixation of the particle number and the spin number, or the spin particle numbers

$$N_\sigma = \int n_\sigma(\vec{r}) d\vec{r}, \qquad \sigma = \uparrow, \downarrow, \tag{9}$$

respectively. The connection between the two representations is given by

$$n_\uparrow(\vec{r}) = \frac{1}{2}(n(\vec{r}) + n_s(\vec{r})) \tag{10a}$$

and

$$n_\downarrow(\vec{r}) = \frac{1}{2}(n(\vec{r}) - n_s(\vec{r})). \tag{10b}$$

For the treatment of a general magnetic field $\vec{B}(\vec{r})$, the involvement of the magnetization density $\vec{m}(\vec{r})$ is necessary in the place of its third component, $-\beta_e n_s(\vec{r})$ [6,7]; however, in this theory, the magnetic field still couples only to the spin of the electrons. To include coupling to orbital currents as well, and to treat diamagnetic effects, a more general extension of DFT is needed – namely, current-density functional theory, either in a non-relativistic or a relativistic form [7,12].

To have unique derivatives in the above Euler-Lagrange equations, a generalization of DFT for noninteger particle and spin numbers is necessary. The cornerstone of such a generalization is the extension of the energy $E[N,v]$, or $E[N,N_s,v,B]$, for fractional values of $N$ and $N_s$. A natural way to achieve this is the zero-temperature grand canonical ensemble definition for $E$ of Perdew et al. [13] (see also [2,14,15]); namely,

$$E[N,v] = \inf_{\hat{\Gamma} \mapsto N} \text{Tr}[\hat{H}_v \hat{\Gamma}]. \tag{11}$$

In the spin-polarized case, this can be given as [16]

$$E[N, N_s, v, B] = \inf_{\hat{\Gamma} \mapsto N, N_s} \text{Tr}\left[\hat{H}_{v,B} \hat{\Gamma}\right], \quad (12)$$

In Eqs.(11) and (12), the infima are searched for under the constraint $\text{Tr}[\hat{N}\hat{\Gamma}] = N$, and under $\text{Tr}[\hat{N}\hat{\Gamma}] = N$ and $\text{Tr}[\hat{N}_s\hat{\Gamma}] = N_s$, respectively, with $\hat{\Gamma}$ denoting general mixed states, $\hat{\Gamma} = \sum_j |\Psi_j\rangle g_j \langle \Psi_j|$, where $|\Psi_j\rangle$ are Fock space vectors, and $g_j \geq 0$ and $\sum_j g_j = 1$. Eq.(12) can be written also with $(N_\uparrow, N_\downarrow)$ in the place of $(N, N_s)$, because of the one-to-one correspondence between $(N, N_s)$ and $(N_\uparrow, N_\downarrow)$. With the above generalization of the energy, DFT and SDFT are naturally extendable for fractional $N$ and $N_s$ via Lieb's Legendre-transform formulation of DFT [2,16]. The Lagrange multipliers $\mu$, $\mu_s$, and $\mu_\sigma$ in Eqs.(4), (7) and (8) can be identified as chemical potentials, namely, as the derivatives of the energy with respect to $N$, $N_s$, and $N_\sigma$, respectively [17,18,13,19]. (Generally, the Lagrange multipliers are derivatives of the minimum of the minimized functional with respect to the constrained quantity they are accounting for.) The energy can be generalized in a temperature dependent way, too, giving a finite-temperature grand canonical ensemble definition for it [20-23], which should yield Eqs.(11) and (12) in the zero-temperature limit.

Provided the ground-state energy $E_{g.s.}(M)$ of systems of integer number ($M$) of electrons is a convex function of the electron number at fixed $v(\vec{r})$ (for which there is experimental, and also numerical, evidence for Coulombic potentials [1,13]), Eq.(11) yields the energy for a general particle number as [13]

$$E[N, v] = (1 - \omega) E[M, v] + \omega E[M + 1, v], \quad (13)$$

with $E[M, v] = E_{g.s.[v]}(M)$ and $E[M+1, v] = E_{g.s.[v]}(M+1)$, and $M$ the integer part of $N$, and $\omega$ the fractional part of $N$ (i.e., $\omega = N - M$). Eq.(13) shows that $E(N)$ is composed of straight-line segments connecting the integer particle-number values. Consequently, there are discontinuities in $E(N)$'s derivative at integer $N$'s. On the basis of Eq.(4), these then imply discontinuities in $\dfrac{\delta F[n]}{\delta n(\vec{r})}$, too. With Eq.(13), the chemical potential $\mu$ in Eq.(4) can be shown to be equal to minus the ionization potential $I$ or the electron affinity $A$ [13], depending on the side the derivative is taken on (for integer $N$); namely,

$$\mu^+ = \left.\frac{\partial E[N, v]}{\partial N}\right|_+ = E[N+1, v] - E[N, v] = -A \quad (14a)$$

and

$$\mu^- = \left.\frac{\partial E[N,v]}{\partial N}\right|_- = E[N,v] - E[N-1,v] = -I \ . \tag{14b}$$

The above result is valid for noninteracting electrons, too. Further, separating the interaction-free part, that is, the single-particle kinetic-energy density functional $T_s[n]$, in $F[n]$ $[= T_s[n] + (F[n] - T_s[n]) \stackrel{def}{=} T_s[n] + E_{xc}[n] + J[n]$, with $J[n]$ the classical Coulomb repulsion part], Eq.(4) can be written as

$$\frac{\delta T_s[n]}{\delta n(\vec{r})} + v_{KS}(\vec{r}) = \mu \ , \tag{15}$$

with

$$v_{KS}(\vec{r}) = \frac{\delta(F[n] - T_s[n])}{\delta n(\vec{r})} + v(\vec{r}) = \frac{\delta E_{xc}[n]}{\delta n(\vec{r})} + \frac{\delta J[n]}{\delta n(\vec{r})} + v(\vec{r}) \ . \tag{16}$$

In Eq.(16), $v_{xc}(\vec{r}) \doteq \frac{\delta E_{xc}[n]}{\delta n(\vec{r})}$ is the so-called exchange-correlation (xc) potential, the only part to be approximated in a Kohn-Sham calculation. From the above,

$$\mu^+ = \left.\frac{\partial E_{s.p.}[N, v_{KS}]}{\partial N}\right|_+ = \sum_{i=1}^{N+1} \varepsilon_i^+ - \sum_{i=1}^{N} \varepsilon_i^+ = \varepsilon_{N+1}^+ \tag{17a}$$

and

$$\mu^- = \left.\frac{\partial E_{s.p.}[N, v_{KS}]}{\partial N}\right|_- = \sum_{i=1}^{N} \varepsilon_i^- - \sum_{i=1}^{N-1} \varepsilon_i^- = \varepsilon_N^- \tag{17b}$$

[24], where $\varepsilon_i^+$ ($\varepsilon_i^-$) are the orbital energies of the one-particle equations

$$-\frac{1}{2}\nabla^2 u_i(\vec{r}) + v_{KS}(\vec{r}) u_i(\vec{r}) = \varepsilon_i u_i(\vec{r}) \ , \qquad i = 1,...,N \ , \tag{18}$$

the Kohn-Sham (KS) equations, with a potential $v_{KS}^+(\vec{r})$ [$v_{KS}^-(\vec{r})$] obtained with the right-[left-] side derivatives taken in Eq.(16). Comparing Eqs.(14) and (17), the highest-occupied orbital energy $\varepsilon_N^-$ of the KS equations with potential $v_{KS}^-(\vec{r})$ is identified with minus the ionization potential $I$ of the interacting electron system, and the lowest-unoccupied orbital energy $\varepsilon_{N+1}^+$ of the KS equations with potential $v_{KS}^+(\vec{r})$ with minus the electron affinity $A$ [13]. Utilizing the result [25,26] (see also [27,28]) that

$$\varepsilon_N - v_{KS}(\infty) = -I \ , \tag{19}$$

the asymptotic value of $v_{KS}^-(\vec{r})$ is obtained as

$$v_{KS}^-(\infty) = 0 \tag{20}$$

(except possibly for the zero-measure nodal surface of the highest occupied KS orbital [29]). This means that the accurate KS potentials, with zero asymptotic value, determined from ab initio densities [30] can be considered as $v_{KS}^-(\vec{r})$, i.e., the left-side derivative of the $E_v[n] - T_s[n]$ that is given by the zero-temperature grand canonical ensemble generalization of the energy.

The spin-polarized case is more complicated. A difficult point in obtaining a spin-polarized version of Eq.(13) is that there are several $M$-electron and $(M+1)$-electron states, which have to be "paired" in some way to obtain proper weighted averages corresponding to the $(M+\omega)$-electron states. Without a generalization of Eq.(13) for SDFT, it is difficult to explore the energy surface $E(N, N_s)$ and to calculate the energy derivatives with respect to $N$ and $N_s$. Relying on their infinite separation approach [15], Yang and coworkers have recently given some insight into what the shape of $E(N, N_s)$ should look like [31], irrespective of the concrete form of the definition of the energy for fractional $N$ and $N_s$, but only in the case of ground states without an external magnetic field. (Note also that since they base their arguments on the spin-independent $E_v[n]$ functional, connected with $E[N, v]$, their conclusions do not directly apply to $E[N, N_s, v, B]$.) Perdew and Sagvolden [32], generalizing their earlier, spin-independent study [33], made an explicit analysis of the special case of the ground-state hydrogen atom plus/minus a fractional number of electrons, exhibiting the discontinuity of the spin-polarized xc potential. Attempts [34,35] have also been made to describe the whole energy surface; however, as will be pointed out, with errors.

In this paper, the proper generalization of Eq.(13) for the spin-polarized case will be presented, and the shape of the $E(N, N_s)$ surface will be revealed, in the case of homogeneous magnetic fields. The left/right-side energy derivatives with respect to $N$, $N_\uparrow$, and $N_\downarrow$ will be shown to give minus the fixed-$N_s$, spin-up, and spin-down ionization potentials/electron affinities, respectively, while the energy derivative with respect to $N_s$ gives (signed) half excitation energies to states with $N_s$ increased (or decreased) by 2. The highest-occupied and lowest-unoccupied Kohn-Sham spin-orbital energies will be identified as minus the corresponding spin-up and spin-down ionization potentials and electron affinities. The excitation energies to the states with $N_s \pm 2$ will be obtained as differences between lowest-unoccupied and opposite-spin highest-occupied spin-orbital energies. Finally, the discontinuities of the spin-polarized KS (or xc) potentials will be exhibited and quantitatively connected in the two representations of SDFT. Throughout the paper, $M$, $M_\uparrow$, and $M_\downarrow$ denote integer electron number, integer spin-up electron number, and integer spin-down electron number, respectively, while $M_s$ denotes integer spin number that is composed of integer spin-up and spin-down electron numbers, i.e., $M_s = M_\uparrow - M_\downarrow$. (We emphasize that $M_s$ is not to be confused with the often similarly denoted total spin, $(N_\uparrow - N_\downarrow)/2$. Note that the total spin could be used in the place of the spin number, but with the use of the spin number, the transformation to $(N_\uparrow, N_\downarrow)$ takes a simpler form.) Further, $E_{l.s.}(M, M_s)$ signifies the energy of the lowest-lying energy-eigenstate with electron number $M$ and spin number $M_s$, in a given $(v(\vec{r}), B(\vec{r}))$, that is also a particle-number ($\hat{N}$) and a spin ($\hat{S}_z$) eigenstate. If $M$ has no relevance in a given situation, simply $E_{l.s.}(M_s)$ will be written.

## II. The energy surface

### A. The fixed, integer particle number cut of $E(N, N_s)$

The energy $E[N, N_s, v, B]$ for fractional spin numbers, at a given integer particle number $M$, is naturally defined as the minimum of $\langle \psi | \hat{H}_{v,B} | \psi \rangle$ over the domain of $M$-particle wavefunctions that give spin number $N_s$,

$$E[M, N_s, v, B] = \min_{\psi_M \mapsto N_s} \langle \psi_M | \hat{H}_{v,B} | \psi_M \rangle . \quad (21)$$

$E[N_\uparrow, N_\downarrow, v, B]$ can be obtained simply by writing $M = N_\uparrow + N_\downarrow$ and $N_s = N_\uparrow - N_\downarrow$ in Eq.(21). The above definition gives just the Li ground-state energy for the state ($N_\uparrow = 1.6, N_\downarrow = 1.4$), with nuclear charge Z=3, e.g., and for any ($N_\uparrow = 2 - \omega, N_\downarrow = 1 + \omega; Z = 3$), with $0 \leq \omega \leq 1$). This linear connection between the $N_s = 1$ and $N_s = -1$ Li ground states is in accordance with the constancy condition of Yang and coworkers [36], obtained on the basis of their infinite separation method [15].

In the case of homogeneous magnetic fields, Eq.(21) reduces to the form

$$E[M, N_s, v, B] = \min_{\{c_{M_s}\}} \left\{ \sum_{M_s = -M}^{M} |c_{M_s}|^2 E_{l.s.}(M, M_s) \middle| \sum_{M_s = -M}^{M} |c_{M_s}|^2 M_s = N_s, \sum_{M_s = -M}^{M} |c_{M_s}|^2 = 1 \right\} . \quad (22)$$

To obtain this expression, expand $\psi_M$ in Eq.(21) into $\hat{H}_{v,B}$'s eigenfunctions $\psi_M^i$: $\psi_M = \sum_i c_i \psi_M^i$. With this, $\langle \psi_M | \hat{H}_{v,B} | \psi_M \rangle = \sum_i |c_i|^2 E_i$, and $E[M, N_s, v, B] = \min_{\{c_i\}} \{ \sum_i |c_i|^2 E_i \,|\, \sum_i |c_i|^2 M_{s;i} = N_s, \ldots \}$. Since the minimum is searched, every term that has $E_i > E_{l.s.}(M_{s;i})$ will have a zero $c_i$, yielding Eq.(22) after a re-indexation of the remaining terms.

If, in the case of homogeneous magnetic fields, $E_{l.s.}(M_s)$, for a fixed integer $N$, is convex with respect to $M_s$, Eq.(21) yields a straight-line connection between the $E_{l.s.}(M_s)$ values. Namely,

$$E(M, M_s \pm \omega_s) = \left(1 - \frac{\omega_s}{2}\right) E(M, M_s) + \frac{\omega_s}{2} E(M, M_s \pm 2), \tag{23a}$$

with

$$0 \leq \omega_s \leq 2. \tag{23b}$$

A function (or functional) $f(x)$ is said to be convex if

$$f((1-\alpha) x_1 + \alpha x_2) \leq (1-\alpha) f(x_1) + \alpha f(x_2) \tag{24a}$$

for $0 < \alpha < 1$. For discrete variables $x$, the above definition is worth giving in the form

$$f(x) \leq \left(1 - \frac{x - x_1}{x_2 - x_1}\right) f(x_1) + \frac{x - x_1}{x_2 - x_1} f(x_2), \tag{24b}$$

for any $x_1 < x < x_2$, by applying the transformation $x = (1-\alpha) x_1 + \alpha x_2$. For a convex $E(M)$, it implies

$$E(M) \leq \frac{1}{2} E(M-1) + \frac{1}{2} E(M+1),$$

i.e.,

$$E(M) - E(M-1) \leq E(M+1) - E(M), \tag{25}$$

or for a convex $E(M_s)$,

$$E(M_s) \leq \frac{1}{2} E(M_s - 2) + \frac{1}{2} E(M_s + 2),$$

i.e.,

$$E(M_s) - E(M_s - 2) \leq E(M_s + 2) - E(M_s). \tag{26}$$

If for the $M_s$'s of an interval $[M_s^1, M_s^2]$, Eq.(24) does not hold for $E_{l.s.}(M_s)$, those energy values are simply left out from the straight-line connection. That is, Eq.(21) yields a convex $E(N_s)$ curve anyway; this $E(N_s)$ is the convex hull (or envelope) of $E_{l.s.}(M_s)$ (following simply from a form like Eq.(22) [2]). Fig. 1 gives an illustration of this with the case of the Nitrogen atom.

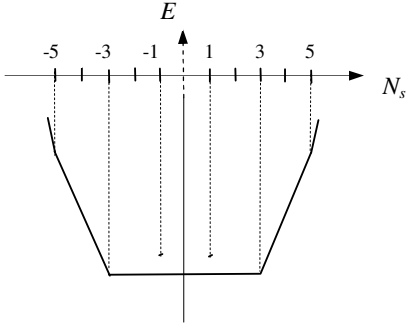

Figure 1. The $E(N_s)$ $(= E[7, N_s, -7/r, 0])$ curve of the Nitrogen atom.

To see the straight-line segment character [Eq.(23)] of $E(N_s)$ given by Eq.(21), consider an $\overline{M}_s$ for which the energy $E_{l.s.}(\overline{M}_s)$ fulfils Eq.(24), i.e.,

$$E_{l.s.}(\overline{M}_s) \leq \left(1 - \frac{\overline{M}_s - M_s^1}{M_s^2 - M_s^1}\right) E_{l.s.}(M_s^1) + \frac{\overline{M}_s - M_s^1}{M_s^2 - M_s^1} E_{l.s.}(M_s^2) \tag{27}$$

for any $M_s^1 < \overline{M}_s < M_s^2$. For $N_s = \overline{M}_s$, the minimum of $\langle \psi_M | \hat{H}_{v,B} | \psi_M \rangle$ will be at the lowest-lying eigenstate $\psi_M^J$ with $M_s^J = N_s$, since any other energy-eigenstate $\psi_M^i$ appearing as a component in $\psi_M \left( = \sum_i \psi_M^i \right)$, with, say, $M_s^i < N_s$, would have to be neutralized by another $\psi_M^i$(s), with $M_s^i > N_s$, for the average of $M_s^i$'s to give $N_s$ – but because of Eq.(27), this would increase $\langle \psi_M | \hat{H}_{v,B} | \psi_M \rangle$. For an $N_s$ that is not an $\overline{M}_s$, the minimum of $\langle \psi_M | \hat{H}_{v,B} | \psi_M \rangle$ will be the corresponding weighted averages of the two nearest $\overline{M}_s$ energy values, $E(N_s) = (1-c) E(\overline{M}_s^1) + c E(\overline{M}_s^2)$ (corresponding to $\psi_M = \sqrt{(1-c)}\, \psi_M^{J_1} + \sqrt{c}\, \psi_M^{J_2}$), with $c = \frac{N_s - \overline{M}_s^1}{\overline{M}_s^2 - \overline{M}_s^1}$. This is so because along any other $\psi_M^i$ appearing in $\psi_M$, with $M_s^i < \overline{M}_s^1$, another $\psi_M^i$(s), with $M_s^i > \overline{M}_s^2$, should appear, which would again increase $\langle \psi_M | \hat{H}_{v,B} | \psi_M \rangle$, due to Eq.(27). Between $\overline{M}_s^1$ and $\overline{M}_s^2$, $E_{l.s.}(M_s) > \left(1 - \frac{M_s - \overline{M}_s^1}{\overline{M}_s^2 - \overline{M}_s^1}\right) E_{l.s.}(\overline{M}_s^1) + \frac{M_s - \overline{M}_s^1}{\overline{M}_s^2 - \overline{M}_s^1} E_{l.s.}(\overline{M}_s^2)$; therefore any

$\psi_M^i$ with $\overline{M}_s^1 < M_s^i < \overline{M}_s^2$ appearing in $\psi_M$ would just increase $\langle \psi_M | \hat{H}_{v,B} | \psi_M \rangle$. In summary, the $E(N_s)$ curve yielded by Eq.(21) can be described by the formula Eq.(23), where $E(M_s)$ (or $E(M_s \pm 2)$) is just the energy $E_{l.s.}(M_s)$ (or $E_{l.s.}(M_s \pm 2)$) if $E_{l.s.}(M_s)$ (or $E_{l.s.}(M_s \pm 2)$) satisfies Eq.(27) (i.e., convexity); otherwise $E(M_s) < E_{l.s.}(M_s)$ (or $E(M_s \pm 2) < E_{l.s.}(M_s \pm 2)$).

## B. The whole surface

Before presenting the formula that describes the energy surface $E(N, N_s)$, it may be useful to consider some possibilities that may be intuitively appealing. In the $(N_\uparrow, N_\downarrow)$ representation of spin-polarized DFT, one would naturally except the $(N_\uparrow = 2, N_\downarrow = 1.5)$ state (with nuclear charge $Z=3$), e.g., to be the 50%-50% mixture of the Li and the Be-like Li$^-$ ground states $(N_\uparrow = 2, N_\downarrow = 1)$ and $(N_\uparrow = 2, N_\downarrow = 2)$. In that case, the energy $E[N_\uparrow, N_\downarrow, v, B]$ could be defined e.g. by the zero-temperature grand canonical ensemble scheme "applied" to the $N_\uparrow$ and $N_\downarrow$ parameters separately:

$$E[M_\uparrow + \omega_\uparrow, N_\downarrow, v, B] = (1 - \omega_\uparrow) E[M_\uparrow, N_\downarrow, v, B] + \omega_\uparrow E[M_\uparrow + 1, N_\downarrow, v, B],$$

which yields

$$E[N_\uparrow, N_\downarrow, v, B] = (1 - \omega_\uparrow)(1 - \omega_\downarrow) E[M_\uparrow, M_\downarrow, v, B] + (1 - \omega_\uparrow) \omega_\downarrow E[M_\uparrow, M_\downarrow + 1, v, B]$$
$$+ \omega_\uparrow (1 - \omega_\downarrow) E[M_\uparrow + 1, M_\downarrow, v, B] + \omega_\uparrow \omega_\downarrow E[M_\uparrow + 1, M_\downarrow + 1, v, B]. \quad (28)$$

However, one should be careful because in this way many states are redefined: not all fractional $N_\sigma$ states correspond to fractional $N$ states (e.g., $N_\uparrow = 1.6$ and $N_\downarrow = 1.4$), which states are therefore already defined (according to Eq.(21)). Writing $\omega_\uparrow = 1 - \omega_\downarrow$ in Eq.(28), the contradiction with Eq.(21) becomes apparent, Eq.(21) giving a constant value for ground states, as pointed out above. Another point against Eq.(28) is the lack of derivative discontinuity with respect to the particle number along the $\omega_\uparrow = \omega_\downarrow$ path. The necessity of a derivative discontinuity at integer $N$'s [13,15,31,32] rules out similarly the $E(N, N_s)$ surface obtained by Vargas et al. [35], since they found no derivative discontinuity along the $N_\uparrow = N_\downarrow$ path between the Li$^+$ and Li$^-$ ground states, e.g. (Of course, it is possible to define $E(N_\uparrow, N_\downarrow)$ for fractional particle numbers in a way that yields no discontinuity at integer $N$'s; however, Janak's functional, used in [35], in itself does not give any fractional $N$ extension – it is a formal framework.)

The $(N, N_s)$ representation has the advantage that it treats the $N$ dependence separately. With the use of it, the energy for a general $N$ may naturally be expected to be defined as

$$E[N, N_s, v, B] = (1-\omega)E[M, N_s, v, B] + \omega E[M+1, N_s, v, B] . \qquad (29)$$

Eq.(29) does better than Eq.(28) in that it yields a derivative discontinuity at integer $N$'s. With Eq.(29), however, the $(N=3.5, N_s=0.5; Z=3)$ state, e.g., is a mixture of the $(N_\uparrow=1.75, N_\downarrow=1.25)$ and $(N_\uparrow=2.25, N_\downarrow=1.75)$ states, and not of the Li and Li$^-$ ground states, contradicting Eq.(21), which gives a straight-line interpolation between $(N=3, N_s=1)$ and $(N=4, N_s=0)$.

Now we turn directly to the zero-temperature grand canonical ensemble definition Eq.(12) to derive the searched energy surface formula. In Eq.(12), the states a statistical mixture $\hat{\Gamma}$ is composed of are not required to have the given $N_s$ separately, but only their average spin number has to give $N_s$. (Of course, constraining the allowed type of $\hat{\Gamma}$'s in Eq.(12) to ones that are composed of states all having the given $N_s$ would be a great help, but is unjustified physically.) Eq.(12) gives a convex surface, similar to Eq.(21) giving a convex curve. If the energy of the lowest-lying spin eigenstates of $\hat{H}_{v,B}$ with particle number $M$ and spin number $M_s$, $E_{l.s.}(M, M_s)$, is convex with respect to $(M, M_s)$, Eq.(12) gives back the corresponding eigenenergies for $(M, M_s)$s, with a straight-line connection between them. Eq.(12) gives Eq.(21) for integer $N$'s and $-N \leq N_s \leq N$, if the energy $E[M, N_s, v, B]$ given by Eq.(21) is convex with respect to $M$. This is simply because Eq.(21) is the convex hull of $E_{l.s.}(M, M_s)$ for a fixed $M$, while Eq.(12) is the convex hull of the whole $E_{l.s.}(M, M_s)$. In this sense, Eq.(21) can be considered as a fixed-$M$ cut of the energy surface. (Of course, mathematically Eq.(12)'s fixed-$M$ cut is nothing else than Eq.(12) with $N=M$, but with this, no additional information is gained.)

To discover the energy surface defined by Eq.(12), it is worth giving it in the form

$$E[N, N_s, v, B] = \inf_{\{p_{MM_s}\}} \left\{ \sum_{M, M_s} p_{MM_s} E_{l.s.}(M, M_s) \Big| \sum_{M, M_s} p_{MM_s} M_s = N_s, \sum_{M, M_s} p_{MM_s} M = N, \sum_{M, M_s} p_{MM_s} = 1 \right\}. \qquad (30)$$

To obtain this form, expand the Fock-space vectors $|\Psi_j\rangle$ of the density operator in Eq.(12) into particle number eigenstates $\psi_M$. With this, $\text{Tr}[\hat{H}_{v,B}\hat{\Gamma}] = \sum_{j,M} p_M^j \langle \psi_M^j | \hat{H}_{v,B} | \psi_M^j \rangle$, $\text{Tr}[\hat{N}_s\hat{\Gamma}] = \sum_{j,M} p_M^j \langle \psi_M^j | \hat{N}_s | \psi_M^j \rangle$, and $\text{Tr}[\hat{N}\hat{\Gamma}] = \sum_{j,M} p_M^j M$. Now, expand $\psi_M^j$ into the eigenstates of

$\hat{H}_{v,B}$ to obtain $\text{Tr}[\hat{H}_{v,B}\hat{\Gamma}] = \sum_{j,M,k} p_M^j |c_{Mk}^j|^2 E_k(M)$ and $\text{Tr}[\hat{N}_s\hat{\Gamma}] = \sum_{j,M,k} p_M^j |c_{Mk}^j|^2 M_{s;Mk}$. Since the infimum of $\text{Tr}[\hat{H}_{v,B}\hat{\Gamma}]$ is taken, terms with $E_k(M) > E_{l.s.}(M, M_{s;Mk})$ will have a zero $c_k$. After a re-indexation in $k$, and introducing $p_{Mk}^j = p_M^j |c_{Mk}^j|^2$, this gives the above form for $E[N, N_s, v, B]$. Note that the index $j$ can be ignored in this expression, since that additional degree of freedom of varying $p_{MM_s}^j$ s does not have any effect in the minimization.

The three-dimensional extension of the straight-line connection of two-dimensional points is a triangular connection formula. The surface yielded by Eq.(12) can be described by the following formula:

$$E(M_\uparrow \pm \omega_\uparrow, M_\downarrow \pm \omega_\downarrow) = (1 - \omega_\uparrow - \omega_\downarrow) E(M_\uparrow, M_\downarrow) + \omega_\uparrow E(M_\uparrow \pm 1, M_\downarrow) + \omega_\downarrow E(M_\uparrow, M_\downarrow \pm 1), \quad (31a)$$

with

$$0 \leq \omega_\uparrow + \omega_\downarrow \leq 1, \quad (31b)$$

where $\omega_\uparrow$ and $\omega_\downarrow$ are nonnegative real numbers, and $M_\uparrow$ and $M_\downarrow$ are integers. (In Eq.(31a), all $\pm$ are in synch.) $E(M_\uparrow, M_\downarrow)$ is just $E_{l.s.}(M_\uparrow, M_\downarrow)$ if $E_{l.s.}(M_\uparrow, M_\downarrow)$ is convex with respect to $(M_\uparrow, M_\downarrow)$ [37] (or equivalently, with respect to $(M, M_s)$ – which is a stricter condition than convexity with respect to $M$ and with respect to $M_s$). Eq.(31a) has been given also by Chan [34], but with an incorrect condition for the omegas; namely, $0 < \omega_\sigma < 1$. Note that Eq.(31b) is crucial to obtain a correct surface; otherwise, a similar error occurs as in the case of [35]. In the $(N, N_s)$ representation (see Fig. 2 for a help), the above formula takes the form

$$E(M \pm \omega, M_s + \omega_s) = (1 - \omega) E(M, M_s) + \frac{\omega + \omega_s}{2} E(M \pm 1, M_s \pm 1) + \frac{\omega - \omega_s}{2} E(M \pm 1, M_s \mp 1), \quad (32a)$$

with

$$0 \leq \omega \leq 1 \quad \text{and} \quad |\omega_s| \leq \omega, \quad (32b)$$

where $M$ and $(M+M_s)/2$ are integer.

It is also useful to give the energy surface formula in a form that describes the triangular elements of the surface from the bottom to the top:

$$E(M \pm \omega, M_s + \omega_s) = \omega E(M \pm 1, M_s + 1) + \left(1 - \omega - \frac{\omega_s}{2}\right) E(M, M_s) + \frac{\omega_s}{2} E(M, M_s + 2), \quad (33a)$$

with

$$0 \leq \omega \leq 1 \quad \text{and} \quad 0 \leq \omega_s \leq 2(1 - \omega). \quad (33b)$$

Eq.(33) is valid also with a minus instead of the plus after the $M_s$'s. It can be observed that, for $\omega = 0$, Eq.(33) reduces to Eq.(23), as is expected.

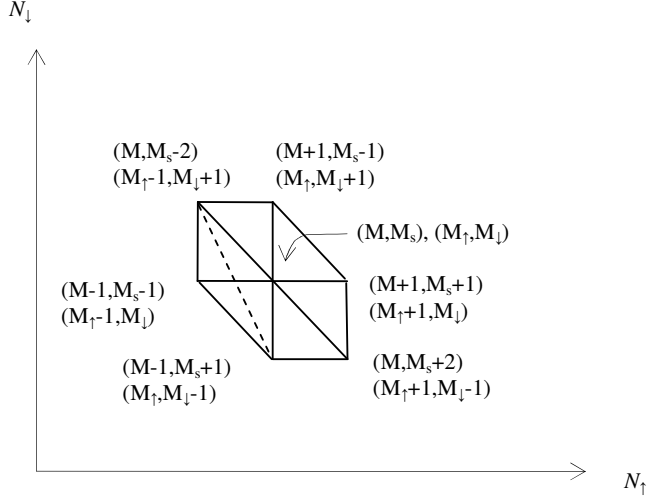

Figure 2. Segment of the $(N_\uparrow, N_\downarrow)$ plane.

If $E_{l.s.}(M, M_s)$ is not convex over an interval $[M_s^1, M_s^2]$, or $[M^1, M^2]$, Eq.(31) (and Eqs.(32) and (33)) is still valid, but $E(M, M_s)$ for those $M$'s and $M_s$'s will not give $E_{l.s.}(M, M_s)$ anymore; it will give a value below that. (Eq.(12) is the convex hull of the lowest-lying $(M, M_s)$ eigenstate energies !) A typical example is a system with an $|N_s| > 1$ ground state in zero external magnetic field, like the Nitrogen atom, or a system with a fully polarized ground state (again with $B(\vec{r}) = 0$), with no derivative discontinuity with respect to the spin number between -3 and 3 (see Fig. 1), and between $-N$ and $N$, respectively. This means that SDFT is unable to deliver the energy of certain lowest-lying $(M, M_s)$ energy-eigenstates (where there is no derivative discontinuity with respect to $N_s$ – like the $E_{l.s.}(N_s = \pm 1)$ state of Nitrogen). For such states, a (real) excited-state theory is needed.

Fig. 3a displays the energy surface yielded by the above formulae, without an external magnetic field. It can be seen that there may be a derivative discontinuity at integer $N$'s, $N_\uparrow$'s and $N_\downarrow$'s.

The only exception to the validity of Eqs.(31)-(33) are cases where the addition of an electron to the ground state, e.g., to get the ground state with $N+1$ electrons goes together with an additional spin flip. As a consequence, the spin number, instead of increasing/decreasing by one, increases/decreases by 3 (or more). In this case, for example, the $E(M, M_s - 2)$ and $E(M, M_s)$ points of the energy surface, instead of being connected by a triangle with $E(M-1, M_s - 1)$, are connected with $E(M-1, M_s + 1)$ (see Fig. 2). The corresponding triangle formula is

$$E(M-\omega, M_s - \omega_s) = \omega E(M-1, M_s + 1) + \left(1 - \omega - \frac{\omega_s}{2}\right) E(M, M_s) + \frac{\omega_s}{2} E(M, M_s - 2)$$

(which can be originated from the $E(M-\omega, M_s - \omega_s)$ version of Eq.(33), by replacing $E(M-1, M_s - 1)$ by $E(M-1, M_s + 1)$ in it), or for the general case of an arbitrary number $k$ of additional spin flips,

$$E(M-\omega, M_s - \omega_s) = \omega E(M-1, M_s + 1) + \left(1 - \omega - \frac{\omega_s}{2}\right) E(M, M_s) + \frac{\omega_s}{2} E(M, M_s - 2k) , \quad (34a)$$

with

$$0 \leq \omega \leq 1 \quad \text{and} \quad 0 \leq \omega_s \leq 2k(1-\omega) . \quad (34b)$$

As an example, the case of certain transition metals, with a $ns^2(n-1)d^3$ electron configuration, can be mentioned. Due to a half-filled d shell being energetically advantageous, the addition of an electron yields a $ns^1(n-1)d^5$ configuration, the spin direction of an $ns$ electron flipping additionally.

Another exception would be if $(1/2)E(M-1, M_s + 1) + (1/2)E(M+1, M_s + 1) < (1/2)E(M, M_s) + (1/2)E(M, M_s + 2)$; that is, if $E(M-1, M_s + 1)$ was connected with $E(M+1, M_s + 1)$ by a straight line, instead of $E(M, M_s)$ being connected with $E(M, M_s + 2)$. This would mean that there is no derivative discontinuity of the energy surface with respect to $N$ at $M$, but there is a derivative discontinuity with respect to $N_s$ at $M_s$. Further, the constancy [36], or generally, linearity, condition on $E(M, N_s)$ between $N_s$ values $M_s$ and $M_s + 2$ would not hold either. (Notice the inherent connection between a derivative discontinuity at $M$ and the linear connection of $E(M_s)$ with $E(M_s + 2)$ at $M$.) However, a surface containing segments like this would, or could, still be convex. By a condition that the energy $E[M, N_s, v, B]$ given by Eq.(21) is convex with respect to $M$ (which is likely in the case of real electronic systems), this case is excluded, and Eq.(31) (or the corresponding formulae for the case of additional spin flips, like Eq.(34)) is the only formula describing the energy surface Eq.(12).

If a homogeneous magnetic field $B$ is switched on (which case is also embraced by Eqs.(31)-(33)), the slope $d$ of the constant-$N$-line segments of the energy surface will change uniformly by $-\beta_e B$; see Fig. 3b. For,

$$\Delta d = \frac{E[M, M_s+2, v, B] - E[M, M_s, v, B]}{2} - \frac{E[M, M_s+2, v, 0] - E[M, M_s, v, 0]}{2}$$

$$= \frac{E[M, M_s+2, v, 0] - (M_s+2)\beta_e B - \left(E[M, M_s, v, 0] - M_s \beta_e B\right)}{2} - \frac{E[M, M_s+2, v, 0] - E[M, M_s, v, 0]}{2} = -\beta_e B. \quad (35)$$

When, for a given particle number $N$, a ground-state level crossing is reached as increasing $B$, the integer-$N$ straight-line segment connecting the ground-state and the (until that moment) first excited-state energy will become horizontal, i.e., constant-energy. (See also the $E(B)$ figure, and the discussion below it, in [19]. In [19], it is shown why the nonuniqueness of external magnetic fields [38] implies a *discontinuity* of the energy derivatives for ground states. Note that since any lowest-lying $N_s$ eigenstate will be a ground state at some $B(\vec{r})$, the derivative discontinuity is implied by $B(\vec{r})$'s nonuniqueness in general.)

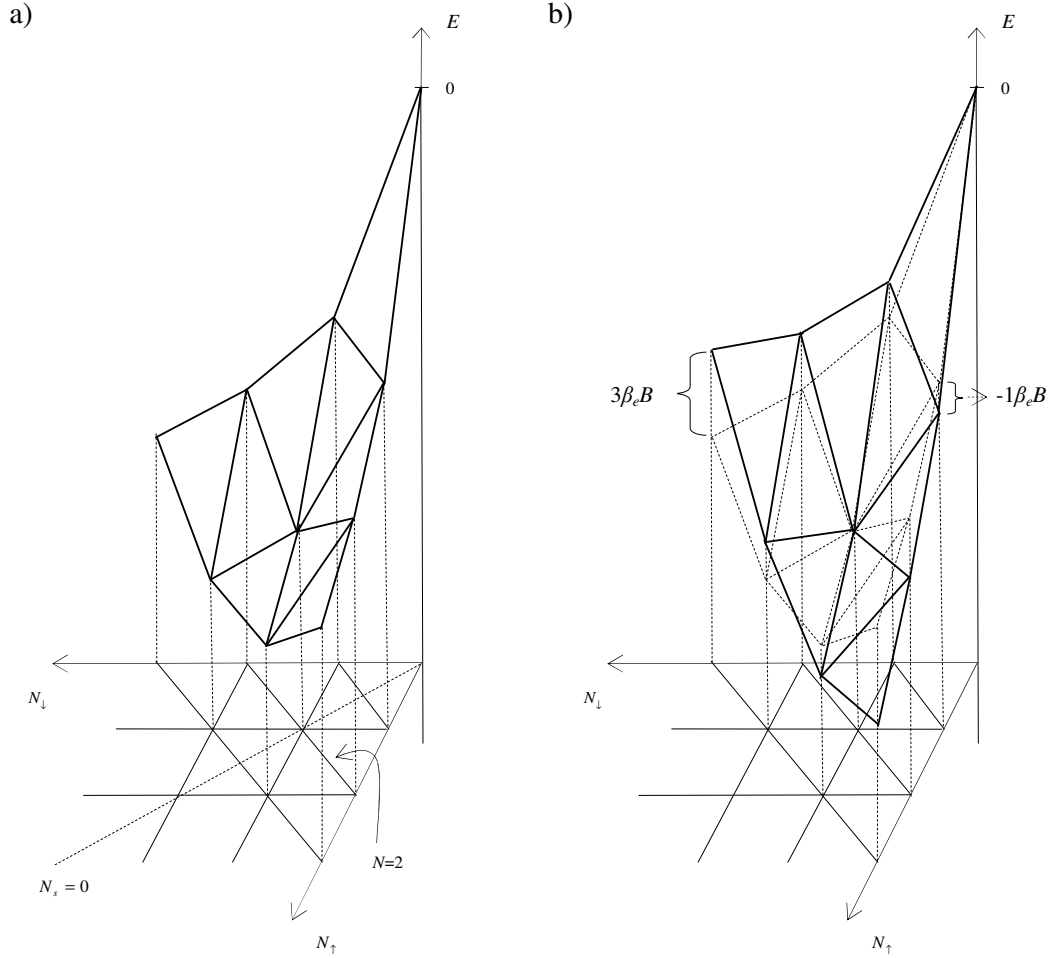

Figure 3. Shape of the energy surface $E(N_\uparrow, N_\downarrow)$ at a given $v(\vec{r})$ (the O atom, e.g., with $v(\vec{r}) = -8/r$), a) without external magnetic field, and b) with a $B(\vec{r}) = const. (> 0)$.

## III. Chemical potentials, Kohn-Sham energies, and derivative discontinuities

### Chemical potentials

With the help of the formulae of Sec.II describing the energy surface $E(N_\uparrow, N_\downarrow)$, the derivatives of the energy with respect to $N$ and $N_s$, or $N_\uparrow$ and $N_\downarrow$, i.e., the spin-resolved chemical potentials, can be easily determined. (The cases where $E_{l.s.}(M, M_s)$ is not convex will be discussed at the end of the next subsection.) In the $(N, N_s)$ representation, with the help of Eq.(32),

$$\mu^+ = \left.\frac{\partial E[N, M_s, v, B]}{\partial N}\right|_+ = \frac{1}{2}\left(E[M+1, M_s+1, v, B] + E[M+1, M_s-1, v, B]\right) - E[M, M_s, v, B]$$

$$= -\frac{1}{2}(A_\uparrow + A_\downarrow) = -A\big|_{N_s} \tag{36a}$$

and

$$\mu^- = \left.\frac{\partial E[N, M_s, v, B]}{\partial N}\right|_- = E[M, M_s, v, B] - \frac{1}{2}\left(E[M-1, M_s+1, v, B] + E[M-1, M_s-1, v, B]\right)$$

$$= -\frac{1}{2}(I_\uparrow + I_\downarrow) = -I\big|_{N_s} \tag{36b}$$

arise for $\mu^{+/-}[N, M_s, v, B]$, with $M$ being the integer part of $N$. (For simplicity in notation, the arguments of $\mu^{+/-}[N, M_s, v, B]$, $A_\sigma[M, M_s, v, B]$, and $I_\sigma[M, M_s, v, B]$ are not displayed.) The chemical potential $\mu$ thus gives (minus) the average of the spin-up and spin-down electron affinities, or ionization potentials, which can be considered also as the electron affinity, or the ionization potential, along constant spin number, respectively. $\mu_s$, on the other hand, yields *signed* half excitation energies. From Eq.(33),

$$\mu_s^+ = \left.\frac{\partial E[M, N_s, v, B]}{\partial N_s}\right|_+ = \frac{1}{2}\left(E[M, M_s+2, v, B] - E[M, M_s, v, B]\right) \tag{37a}$$

and

$$\mu_s^- = \left.\frac{\partial E[M, N_s, v, B]}{\partial N_s}\right|_- = \frac{1}{2}\left(E[M, M_s, v, B] - E[M, M_s-2, v, B]\right). \tag{37b}$$

As can be seen, $\mu_s^+$ gives half of the (minimum) energy "needed" to turn a spin-down electron into a spin-up electron, while $\mu_s^-$ gives half of the energy "gained" by turning a spin-up electron into a spin-down one.

In the $(N_\uparrow, N_\downarrow)$ representation, the spin-resolved chemical potentials are obtained from Eq.(31) as

$$\mu_\uparrow^+ = \left.\frac{\partial E[N_\uparrow, M_\downarrow, v, B]}{\partial N_\uparrow}\right|_+ = E[M_\uparrow+1, M_\downarrow, v, B] - E[M_\uparrow, M_\downarrow, v, B] = -A_\uparrow, \tag{38a}$$

$$\mu_\uparrow^- = \left.\frac{\partial E[N_\uparrow, M_\downarrow, v, B]}{\partial N_\uparrow}\right|_- = E[M_\uparrow, M_\downarrow, v, B] - E[M_\uparrow-1, M_\downarrow, v, B] = -I_\uparrow, \tag{38b}$$

$$\mu_\downarrow^+ = \left.\frac{\partial E[M_\uparrow, N_\downarrow, v, B]}{\partial N_\downarrow}\right|_+ = E[M_\uparrow, M_\downarrow+1, v, B] - E[M_\uparrow, M_\downarrow, v, B] = -A_\downarrow, \tag{39a}$$

and

$$\mu_\downarrow^- = \left.\frac{\partial E[M_\uparrow, N_\downarrow, v, B]}{\partial N_\downarrow}\right|_- = E[M_\uparrow, M_\downarrow, v, B] - E[M_\uparrow, M_\downarrow - 1, v, B] = -I_\downarrow . \quad (39b)$$

$\mu_\uparrow^-$ ($\mu_\downarrow^-$) thus gives the negative of the minimum energy needed to remove a spin-up (spin-down) electron, while $\mu_\uparrow^+$ ($\mu_\downarrow^+$) gives the negative of the maximum energy gained by adding a spin-up (spin-down) electron to the $(M_\uparrow, M_\downarrow)$-electron system.

In Eqs.(36)-(39), the chemical potentials are calculated at $N_s = M_s$, $N = M$, $N_\downarrow = M_\downarrow$, and $N_\uparrow = M_\uparrow$, respectively, but with the help of Eqs.(31)-(33), they can easily be determined at other points, too.

## Highest-occupied and lowest-unoccupied Kohn-Sham spin energies and energy differences

Separating the single-particle kinetic energy functional $T_s[n_\uparrow, n_\downarrow]$ in $F[n_\uparrow, n_\downarrow]$ yields the SDFT Euler-Lagrange equations Eq.(8) in the form

$$\frac{\delta T_s[n_\uparrow, n_\downarrow]}{\delta n_\uparrow(\vec{r})} + v_{KS}^\uparrow(\vec{r}) = \mu_\uparrow \quad (40a)$$

and

$$\frac{\delta T_s[n_\uparrow, n_\downarrow]}{\delta n_\downarrow(\vec{r})} + v_{KS}^\downarrow(\vec{r}) = \mu_\downarrow , \quad (40b)$$

with

$$v_{KS}^\sigma(\vec{r}) = \frac{\delta(F[n_\uparrow, n_\downarrow] - T_s[n_\uparrow, n_\downarrow])}{\delta n_\sigma(\vec{r})} + v(\vec{r}) - \tau_z^{\sigma\sigma} \beta_e B(\vec{r}) , \quad (41)$$

where $\tau_z$ denotes the third Pauli matrix (with $\tau_z^{\uparrow\uparrow} = 1$ and $\tau_z^{\downarrow\downarrow} = -1$). The above equations are the SDFT Euler-Lagrange equations for a system of noninteracting identical fermions with the given $N$ and $N_s$ in external electrostatic and (collinear) magnetic fields $\frac{1}{2}\left(v_{KS}^\uparrow(\vec{r}) + v_{KS}^\downarrow(\vec{r})\right)$ and $-\frac{1}{2\beta_e}\left(v_{KS}^\uparrow(\vec{r}) - v_{KS}^\downarrow(\vec{r})\right)$. Such a system is described by the single-particle Schrödinger equations

$$-\frac{1}{2}\nabla^2 u_i^\uparrow(\vec{r}) + v_{KS}^\uparrow(\vec{r}) u_i^\uparrow(\vec{r}) = \varepsilon_i^\uparrow u_i^\uparrow(\vec{r}) , \quad i = 1,...,N_\uparrow , \quad (42a)$$

$$-\frac{1}{2}\nabla^2 u_i^\downarrow(\vec{r}) + v_{KS}^\downarrow(\vec{r}) u_i^\downarrow(\vec{r}) = \varepsilon_i^\downarrow u_i^\downarrow(\vec{r}) , \quad i = 1,...,N_\downarrow , \quad (42b)$$

the spin-polarized Kohn-Sham equations. Consequently, $\mu_\uparrow$ and $\mu_\downarrow$ emerge on the basis of Eqs.(38) and (39), with $E_{s.p.}[N_\uparrow, N_\downarrow, \frac{1}{2}(v_{KS}^\uparrow(\vec{r}) + v_{KS}^\downarrow(\vec{r})), -\frac{1}{2\beta_e}(v_{KS}^\uparrow(\vec{r}) - v_{KS}^\downarrow(\vec{r}))]$, as

$$\mu_\uparrow^+ = \left(\sum_{i=1}^{N_\uparrow+1} \varepsilon_i^{\uparrow,+} + \sum_{i=1}^{N_\downarrow} \varepsilon_i^\downarrow \right) - \left(\sum_{i=1}^{N_\uparrow} \varepsilon_i^{\uparrow,+} + \sum_{i=1}^{N_\downarrow} \varepsilon_i^\downarrow \right) = \varepsilon_{N_\uparrow+1}^{\uparrow,+}, \quad (43a)$$

$$\mu_\uparrow^- = \left(\sum_{i=1}^{N_\uparrow} \varepsilon_i^{\uparrow,-} + \sum_{i=1}^{N_\downarrow} \varepsilon_i^\downarrow \right) - \left(\sum_{i=1}^{N_\uparrow-1} \varepsilon_i^{\uparrow,-} + \sum_{i=1}^{N_\downarrow} \varepsilon_i^\downarrow \right) = \varepsilon_{N_\uparrow}^{\uparrow,-}, \quad (43b)$$

and

$$\mu_\downarrow^+ = \left(\sum_{i=1}^{N_\uparrow} \varepsilon_i^\uparrow + \sum_{i=1}^{N_\downarrow+1} \varepsilon_i^{\downarrow,+} \right) - \left(\sum_{i=1}^{N_\uparrow} \varepsilon_i^\uparrow + \sum_{i=1}^{N_\downarrow} \varepsilon_i^{\downarrow,+} \right) = \varepsilon_{N_\downarrow+1}^{\downarrow,+}, \quad (44a)$$

$$\mu_\downarrow^- = \left(\sum_{i=1}^{N_\uparrow} \varepsilon_i^\uparrow + \sum_{i=1}^{N_\downarrow} \varepsilon_i^{\downarrow,-} \right) - \left(\sum_{i=1}^{N_\uparrow} \varepsilon_i^\uparrow + \sum_{i=1}^{N_\downarrow-1} \varepsilon_i^{\downarrow,-} \right) = \varepsilon_{N_\downarrow}^{\downarrow,-}. \quad (44b)$$

In Eqs.(43) and (44), $\varepsilon_i^{\sigma,+}$ ($\varepsilon_i^{\sigma,-}$) denotes the KS energies obtained with the right- (left-) side KS potential $v_{KS}^{\sigma,+}(\vec{r})$ ($v_{KS}^{\sigma,-}(\vec{r})$) [i.e., with right- (left-) side derivatives in Eqs.(40) and (41)] inserted into the KS equations [Eq.(42)]. (There being no + or – in the superscript means a free choice of the asymptotic value of the corresponding KS potential.) The electrostatic part of the KS potential, $\frac{1}{2}\left(v_{KS}^\uparrow(\vec{r}) + v_{KS}^\downarrow(\vec{r})\right)$, does not necessarily go to zero at infinity, contrary to a "real" potential $v(\vec{r})$; however, this does not cause any problem in Eqs.(43) and (44), since a shift of the electrostatic potential does not alter the relative positions of the single-particle energies. The only criterion in obtaining Eqs.(38) and (39), and Eqs.(36) and (37), was the convexity of $E(M, M_s)$, which holds for any noninteracting-electron system, irrespective of the asymptotic value of the electrostatic potential (see Appendix A). It has to be mentioned here that for not noninteracting-($v$,$B$)-representable $(n_\uparrow(\vec{r}), n_\downarrow(\vec{r}))$s (or $(n(\vec{r}), n_s(\vec{r}))$s) [39], with hole(s) below the spin-HOMOs, or with a nonexisting $T_s[n_\uparrow, n_\downarrow]$ derivative, the above (and following) identification of the chemical potentials as KS energies (or energy differences, below) breaks down, just as in the spin-independent case.

Since the chemical potential of the KS system equals the chemical potential of the corresponding interacting system by construction (see Appendix B), by confronting Eqs.(43) and (44) with Eqs.(38) and (39), the highest occupied KS spin-orbital energies $\varepsilon_{N_\uparrow}^\uparrow$ and $\varepsilon_{N_\downarrow}^\downarrow$ are identified as the negatives of the spin-up and spin-down ionization potentials of the $(N_\uparrow, N_\downarrow)$–electron system, respectively, and the lowest unoccupied KS spin-orbital energies $\varepsilon^{\uparrow}_{N_\uparrow+1}$ and $\varepsilon^{\downarrow}_{N_\downarrow+1}$ are identified as the negatives of the spin-up and spin-down electron affinities; i.e.,

$$\varepsilon^{\sigma,+}_{N_\sigma+1} = -A_\sigma \qquad (45a)$$

and

$$\varepsilon^{\sigma,-}_{N_\sigma} = -I_\sigma . \qquad (45b)$$

It is worth noting that in the derivation of Eqs.(45), no explicit involvement of fractional occupations has been needed. Eqs.(45) intuitively are a straightforward generalization of the spin-independent result, Eq.(17). Chan [34] has also obtained them, together with Eqs.(38) and (39); however, without a correct analytical description of $E(N, N_s)$, they were not rigorously established.

In the $(N, N_s)$ representation, separating $T_s[n, n_s]$ in $F[n, n_s]$ yields the SDFT Euler-Lagrange equations Eq.(7) in the form

$$\frac{\delta T_s[n, n_s]}{\delta n(\vec{r})} + v_{KS}(\vec{r}) = \mu \qquad (46a)$$

and

$$\frac{\delta T_s[n, n_s]}{\delta n_s(\vec{r})} + v^s_{KS}(\vec{r}) = \mu_s , \qquad (46b)$$

with

$$v_{KS}(\vec{r}) = \frac{\delta(F[n, n_s] - T_s[n, n_s])}{\delta n(\vec{r})} + v(\vec{r}) \qquad (47a)$$

and

$$v^s_{KS}(\vec{r}) = \frac{\delta(F[n, n_s] - T_s[n, n_s])}{\delta n_s(\vec{r})} - \beta_e B(\vec{r}) . \qquad (47b)$$

(This $v_{KS}(\vec{r})[n, n_s]$, of course, is not the same as $v_{KS}(\vec{r})[n]$ of spin-independent DFT.) The corresponding single-particle equations are

$$-\frac{1}{2}\nabla^2 u^{\uparrow}_i(\vec{r}) + \left(v_{KS}(\vec{r}) + v^s_{KS}(\vec{r})\right)u^{\uparrow}_i(\vec{r}) = \varepsilon^{\uparrow,s}_i u^{\uparrow}_i(\vec{r}) , \quad i = 1,...,N_\uparrow , \qquad (48a)$$

$$-\frac{1}{2}\nabla^2 u^{\downarrow}_i(\vec{r}) + \left(v_{KS}(\vec{r}) - v^s_{KS}(\vec{r})\right)u^{\downarrow}_i(\vec{r}) = \varepsilon^{\downarrow,s}_i u^{\downarrow}_i(\vec{r}) , \quad i = 1,...,N_\downarrow .\qquad (48b)$$

The chemical potentials $\mu$ and $\mu_s$ can now be given on the basis of Eqs.(36) and (37) as

$$\mu^+ = \frac{1}{2}\left(\varepsilon^{\uparrow,s(+/-)}_{N_\uparrow+1} + v^+_{KS}(\infty)\right) + \frac{1}{2}\left(\varepsilon^{\downarrow,s(+/-)}_{N_\downarrow+1} + v^+_{KS}(\infty)\right)$$

$$= \frac{1}{2}\left(\varepsilon^{\uparrow,s(+/-)}_{N_\uparrow+1} + \varepsilon^{\downarrow,s(+/-)}_{N_\downarrow+1}\right) + \frac{1}{2}\left(v^{\uparrow,+}_{KS}(\infty) + v^{\downarrow,+}_{KS}(\infty)\right) = \frac{1}{2}\left(\varepsilon^{\uparrow,+}_{N_\uparrow+1} + \varepsilon^{\downarrow,+}_{N_\downarrow+1}\right), \qquad (49a)$$

$$\mu^- = \frac{1}{2}\varepsilon^{\uparrow,s(+/-)}_{N_\uparrow} + \frac{1}{2}\varepsilon^{\downarrow,s(+/-)}_{N_\downarrow}$$

$$= \frac{1}{2}\left(\varepsilon^{\uparrow,s(+/-)}_{N_\uparrow} + \varepsilon^{\downarrow,s(+/-)}_{N_\downarrow}\right) + \frac{1}{2}\left(v^{\uparrow,-}_{KS}(\infty) + v^{\downarrow,-}_{KS}(\infty)\right) = \frac{1}{2}\left(\varepsilon^{\uparrow,-}_{N_\uparrow} + \varepsilon^{\downarrow,-}_{N_\downarrow}\right), \qquad (49b)$$

and

$$\mu^+_s = \frac{1}{2}\left\{\left(\sum_{i=1}^{N_\uparrow+1}\varepsilon^{\uparrow,s+}_i + \sum_{i=1}^{N_\downarrow-1}\varepsilon^{\downarrow,s+}_i\right) - \left(\sum_{i=1}^{N_\uparrow}\varepsilon^{\uparrow,s+}_i + \sum_{i=1}^{N_\downarrow}\varepsilon^{\downarrow,s+}_i\right)\right\} = \frac{1}{2}\left(\varepsilon^{\uparrow,s+}_{N_\uparrow+1} - \varepsilon^{\downarrow,s+}_{N_\downarrow}\right), \qquad (50a)$$

$$\mu^-_s = \frac{1}{2}\left\{\left(\sum_{i=1}^{N_\uparrow}\varepsilon^{\uparrow,s-}_i + \sum_{i=1}^{N_\downarrow}\varepsilon^{\downarrow,s-}_i\right) - \left(\sum_{i=1}^{N_\uparrow-1}\varepsilon^{\uparrow,s-}_i + \sum_{i=1}^{N_\downarrow+1}\varepsilon^{\downarrow,s-}_i\right)\right\} = \frac{1}{2}\left(\varepsilon^{\uparrow,s-}_{N_\uparrow} - \varepsilon^{\downarrow,s-}_{N_\downarrow+1}\right). \qquad (50b)$$

Above, $\varepsilon^{\sigma,s+}_i$ ($\varepsilon^{\sigma,s-}_i$) denotes the KS energies obtained with the KS potential $v^-_{KS}(\vec{r}) + \tau^{\sigma\sigma}_z v^{s,+}_{KS}(\vec{r})$ [$v^-_{KS}(\vec{r}) + \tau^{\sigma\sigma}_z v^{s,-}_{KS}(\vec{r})$] inserted into the KS equations. (The asymptotic constant of $v_{KS}(\vec{r})$, as $v^+_{KS}(\infty)$ or $v^-_{KS}(\infty)$, can be chosen freely, since it cancels out in Eqs.(50). The equalities $v^\pm_{KS}(\infty) = \frac{1}{2}\left(v^{\uparrow,\pm}_{KS}(\infty) + v^{\downarrow,\pm}_{KS}(\infty)\right)$, used in Eqs.(49), will be derived later; see Eqs.(62) and (63).)

Eqs.(49) and (50) determine the gaps in $F[n,n_s]$'s derivative in terms of KS energies. Comparing with Eq.(37), Eq.(50) gives the excitation energies to states with $N_s (= M_s)$ increased/decreased by 2 via KS energies in an intuitively appealing way,

$$\varepsilon^{\uparrow,s+}_{M_\uparrow+1} - \varepsilon^{\downarrow,s+}_{M_\downarrow} = E[M, M_s+2, v, B] - E[M, M_s, v, B], \qquad (51a)$$

$$\varepsilon^{\uparrow,s-}_{M_\uparrow} - \varepsilon^{\downarrow,s-}_{M_\downarrow+1} = E[M, M_s, v, B] - E[M, M_s-2, v, B], \qquad (51b)$$

with $M_\uparrow = (M+M_s)/2$ and $M_\downarrow = (M-M_s)/2$. (Note the importance of the $s$ index of the single-particle energies, implying a shift of the spin-up and spin-down energies relative to each other, as compared to $\varepsilon^\uparrow_i$ and $\varepsilon^\downarrow_i$; see Eq.(70) of the next subsection.) Thus, provided one has an accurate approximation for $E_{xc}[n_\uparrow, n_\downarrow]$ ($= E_{xc}[n,n_s]$) that properly incorporates the derivative discontinuities, from one KS calculation one can obtain the energies of 7 states, namely, the energies of the $(M, M_s)$, $(M, M_s \pm 2)$, $(M+1, M_s \pm 1)$, $(M-1, M_s \pm 1)$ lowest-lying energy eigenstates. (Obtaining the density and the spin density from a KS calculation, using a given pair of xc energy derivatives, then all the necessary xc derivative discontinuities can be determined from $E_{xc}[n_\uparrow, n_\downarrow]$.) Of course, it is possible that one has an approximate $E_{xc}[n_\uparrow, n_\downarrow]$ that accounts for the discontinuities only along $n_\sigma(\vec{r})$'s (but not along $n_s(\vec{r})$), e.g.; in that case, the excitation energies will not be obtained with one calculation.

Besides leading to real excitation energies via Eq.(51), the use of the KS spin-potentials of the $(N, N_s)$ representation, $\left(v_{KS}(\vec{r}) + v^s_{KS}(\vec{r})\right)$ and $\left(v_{KS}(\vec{r}) - v^s_{KS}(\vec{r})\right)$, has another advantageous consequence: It restores the Aufbau principle, observed to break down (for the Li atom) in the case of KS spin-potentials with zero asymptotic value [40]. For Li (with $N_s = -1$), without external magnetic field, $\mu_s^+ (= E[M=3, M_s=1] - E[M=3, M_s=-1]) = 0$, which gives $\varepsilon_{M_\uparrow+1}^{\uparrow,s+} = \varepsilon_{M_\downarrow}^{\downarrow,s+}$, in the place of $\varepsilon_{M_\uparrow+1}^{\uparrow,-} < \varepsilon_{M_\downarrow}^{\downarrow,-}$, corresponding to $v_{KS}^{\uparrow,-}(\vec{r})$ and $v_{KS}^{\downarrow,-}(\vec{r})$, with $v_{KS}^{\uparrow(\downarrow),-}(\infty) = 0$. That is, with $\left(v_{KS}^-(\vec{r}) + v_{KS}^{s,+}(\vec{r})\right)$ and $\left(v_{KS}^-(\vec{r}) - v_{KS}^{s,+}(\vec{r})\right)$, the Aufbau is restored. In the case of $N_s = 1$, these spin-potentials should be replaced by $\left(v_{KS}^-(\vec{r}) + v_{KS}^{s,-}(\vec{r})\right)$ and $\left(v_{KS}^-(\vec{r}) - v_{KS}^{s,-}(\vec{r})\right)$. These considerations can be applied generally for ground-state, open-shell systems, with $N_s \neq 0$, without an external magnetic field. In the case of closed-shell systems (with no magnetic field), where $\varepsilon_{M_\uparrow}^{\uparrow,-} = \varepsilon_{M_\downarrow}^{\downarrow,-}$ (i.e., the Aufbau is valid), now it has to be examined whether shifting $v_{KS}^{\uparrow,-}(\vec{r})$ and $v_{KS}^{\downarrow,-}(\vec{r})$ to $\left(v_{KS}^-(\vec{r}) + v_{KS}^{s,+/-}(\vec{r})\right)$ and $\left(v_{KS}^-(\vec{r}) - v_{KS}^{s,+/-}(\vec{r})\right)$, respectively, ruins the Aufbau principle. Of course, $E(M, M_s + 2) > E(M, M_s)$ (and $E(M, M_s - 2) > E(M, M_s)$) for the systems in question, hence $\varepsilon_{M_\uparrow+1}^{\uparrow,s+} > \varepsilon_{M_\downarrow}^{\downarrow,s+}$ (and $\varepsilon_{M_\downarrow+1}^{\downarrow,s-} > \varepsilon_{M_\uparrow}^{\uparrow,s-}$); but the question is whether $\varepsilon_{M_\uparrow}^{\uparrow,s+} < \varepsilon_{M_\downarrow+1}^{\downarrow,s+}$ ($\varepsilon_{M_\downarrow}^{\downarrow,s-} < \varepsilon_{M_\uparrow+1}^{\uparrow,s-}$). To prove $\varepsilon_{M_\uparrow}^{\uparrow,s+} < \varepsilon_{M_\downarrow+1}^{\downarrow,s+}$, we start from the trivial $\varepsilon_{M_\downarrow}^{\downarrow,-} < \varepsilon_{M_\downarrow+1}^{\downarrow,-}$. Since $\varepsilon_{M_\downarrow}^{\downarrow,-} = \varepsilon_{M_\uparrow}^{\uparrow,-}$ for closed-shell systems with no magnetic field (where $v_{KS}^{\uparrow,-}(\vec{r}) = v_{KS}^{\downarrow,-}(\vec{r})$, and $M_\uparrow = M_\downarrow$), we have $\varepsilon_{M_\uparrow}^{\uparrow,-} < \varepsilon_{M_\downarrow+1}^{\downarrow,-}$, i.e., $\varepsilon_{M_\uparrow}^{\uparrow,-} - \varepsilon_{M_\downarrow+1}^{\downarrow,-} < 0$. Now, we utilize the result of Savin et al. [27] that for closed-shell systems, the difference between the LUMO and the HOMO spin-unresolved KS energy gives a value that is between the energies of the triplet and the singlet excitation from the HOMO; consequently, $E(M, M_s + 2) - E(M, M_s) < \varepsilon_{M/2+1}^- - \varepsilon_{M/2}^-$. From this, since the KS potential (with zero asymptotic value) of spin-independent DFT equals $v_{KS}^{\uparrow(\downarrow),-}(\vec{r})$ for the considered case, $E(M, M_s + 2) - E(M, M_s) < \varepsilon_{M_\uparrow+1}^{\uparrow,-} - \varepsilon_{M_\downarrow}^{\downarrow,-}$ emerges. This has the consequence that $v_{KS}^{s,+}(\infty) < 0$, on the basis of Eq.(69a). Adding $2v_{KS}^{s,+}(\infty) < 0$ to the above $\varepsilon_{M_\uparrow}^{\uparrow,-} - \varepsilon_{M_\downarrow+1}^{\downarrow,-} < 0$, and applying $\varepsilon_{M_\uparrow}^{\uparrow,s+} = \varepsilon_{M_\uparrow}^{\uparrow,-} + v_{KS}^{s,+}(\infty)$ and $\varepsilon_{M_\downarrow+1}^{\downarrow,s+} = \varepsilon_{M_\downarrow+1}^{\downarrow,-} - v_{KS}^{s,+}(\infty)$, we finally have $\varepsilon_{M_\uparrow}^{\uparrow,s+} < \varepsilon_{M_\downarrow+1}^{\downarrow,s+}$. ($\varepsilon_{M_\downarrow}^{\downarrow,s-} < \varepsilon_{M_\uparrow+1}^{\uparrow,s-}$ can be proved similarly.) For completeness, we mention that in the case of open-shell systems with $N_s < 0$ [or $N_s > 0$], the Aufbau principle remains valid with $\left(v_{KS}^-(\vec{r}) + v_{KS}^{s,-}(\vec{r})\right)$ and $\left(v_{KS}^-(\vec{r}) - v_{KS}^{s,-}(\vec{r})\right)$ [or $\left(v_{KS}^-(\vec{r}) + v_{KS}^{s,+}(\vec{r})\right)$ and $\left(v_{KS}^-(\vec{r}) - v_{KS}^{s,+}(\vec{r})\right)$], too, on the basis of similar arguments as these, if the result of [27] is generalizable for open-shell systems.

If there is a missing derivative discontinuity of the energy along $N_s$, the relation with excitation energies modifies. $\mu_s$ (times $\Delta M_s$) will give the excitation energy not to the $(M_s \pm 2)$ energy-eigenstate, but to the next $(M_s + \Delta M_s)$ energy-eigenstate where there is no missing discontinuity. Since $E_{l.s.}^{s.p.}(M,M_s)$ of noninteracting systems is always convex in $M_s$ (see Appendix A), Eq.(50) never breaks down because of a missing $N_s$-derivative discontinuity of $E^{s.p.}(N,N_s)$, defined by Eq.(12) with a single-particle Hamiltonian (the lowest-energy state with an $M_s$ will always be an energy eigenstate for noninteracting systems, i.e., $E^{s.p.}(M,M_s) = E_{l.s.}^{s.p.}(M,M_s)$, contrary to the situation sketched in Fig.1; consequently, though there are no $N_s$-derivative discontinuities of the energy Eq.(12) at certain $M_s$'s for noninteracting systems either, they are not "missing" discontinuities). In the case of the ground-state Nitrogen atom (with, say, $N_s = -3$) without external magnetic field, e.g., $\mu_s^+$ is zero (see Fig.1); that is, Eq.(50) says that $\varepsilon_{N_\uparrow+1}^{\uparrow,s+}$ equals $\varepsilon_{N_\downarrow}^{\downarrow,s+}$. If a homogeneous magnetic field $B$ ($<0$) is switched on, $\varepsilon_{N_\uparrow+1}^{\uparrow,s+}$ and $\varepsilon_{N_\downarrow}^{\downarrow,s+}$ will shift relative to each other. The connection of $\varepsilon_{N_\uparrow+1}^{\uparrow,s+} - \varepsilon_{N_\downarrow}^{\downarrow,s+}$ with excitation energies will be $\varepsilon_{N_\uparrow+1}^{\uparrow,s+} - \varepsilon_{N_\downarrow}^{\downarrow,s+} = \frac{1}{3}(E(N,N_s+6) - E(N,N_s))$; that is, turning all 3 spin-down p electrons of the KS noninteracting system into the opposite direction will give the excitation energy of Nitrogen from its $N_s = -3$ ground state to its $N_s = 3$ excited state, which is quite appealing intuitively. The shift of energy levels due to $B$ can be given explicitly as $E(N,N_s+6) - E(N,N_s) = -6\beta_e B$ and $\varepsilon_{N_\uparrow+1}^{\uparrow,s+} - \varepsilon_{N_\downarrow}^{\downarrow,s+} = -2\beta_e B$ (obtained from Eqs.(47) and (48), with the orbitals remaining unchanged as increasing $-B$), confirming the above relation.

In contrast to the above, in the cases described in the paragraph containing Eq.(34), it is $\mu_\uparrow$ (or $\mu_\downarrow$) what modifies. In the example displayed in Fig. 2, $\mu_\uparrow^-(M_\uparrow, M_\downarrow)$ will give

$$\mu_\uparrow^- = 2E(M_\uparrow, M_\downarrow) - \left(E(M_\uparrow, M_\downarrow - 1) + E(M_\uparrow - 1, M_\downarrow + 1)\right) = -I_\downarrow + E(M_\uparrow, M_\downarrow) - E(M_\uparrow - 1, M_\downarrow + 1) \quad (52)$$

instead of $-I_\uparrow$, for instance.

Having the negatives of the highest occupied KS spin-orbital energies identified with spin-resolved ionization potentials, the asymptotic value of the left-side Kohn-Sham spin-potentials can be shown to be zero, similar to the spin-independent case [25,26]. Gritsenko and Baerends [41] (see also [40]) have obtained the following result for $v_{KS}^\sigma(\infty)$:

$$-I_\sigma = \varepsilon_{N_\sigma}^\sigma - v_{KS}^\sigma(\infty) \ , \tag{53}$$

i.e., the highest-occupied spin-up (spin-down) KS energy calculated with an effective potential vanishing at infinity equals minus the spin-up (spin-down) ionization potential. (Note that Eq.(53) can be generalized to obtain an approximate, but rather accurate, DFT analog of Koopmans' theorem [41,40].) Comparing Eq.(53) with Eq.(45b), one immediatelly finds

$$v_{KS}^{\sigma,-}(\infty) = 0 \ . \tag{54}$$

We emphasize here the formal nature of Eqs.(45a) and (51) in the sense that without knowing the asymptotic values of the KS spin-potentials relative to which the KS energies in those equations are to be measured, Eqs.(45a) and (51) are of no (direct) practical use. But we note also that similarly to the spin-independent case (where on the basis of $\varepsilon_{N+1}^+ = -A$, it is possible to calculate the electron affinity from known densities $n(\vec{r})[N,v]$ and $n(\vec{r})[N+1,v]$), Eqs.(45a) and (51) make it possible to obtain $-A_\sigma$ [34] and $E(M, M_s \pm 2) - E(M, M_s)$, provided one has a knowledge of the corresponding densities in a given $v(\vec{r})$ and $B$.

**Derivative discontinuities**

Accounting for the derivative discontinuities of the energy surface is essential. This shows well in transforming from the $(N_\uparrow, N_\downarrow)$ representation to the $(N, N_s)$ representation. The energy $E(N, N_s)$ can be obtained from $E(N_\uparrow, N_\downarrow)$ via the transformation

$$E[N, N_s, v, B] = E[N_\uparrow = (N+N_s)/2, N_\downarrow = (N-N_s)/2, v, B] \ . \tag{55}$$

Differentiating the above relation with respect to $N$ and $N_s$ yields

$$\left.\frac{\partial E[N, N_s, v, B]}{\partial N}\right|_+ = \frac{1}{2}\left(\left.\frac{\partial E[N_\uparrow, N_\downarrow, v, B]}{\partial N_\uparrow}\right|_+ + \left.\frac{\partial E[N_\uparrow, N_\downarrow, v, B]}{\partial N_\downarrow}\right|_+\right) \tag{56}$$

and

$$\left.\frac{\partial E[N, N_s, v, B]}{\partial N_s}\right|_+ = \frac{1}{2}\left(\left.\frac{\partial E[N_\uparrow, N_\downarrow, v, B]}{\partial N_\uparrow}\right|_+ - \left.\frac{\partial E[N_\uparrow, N_\downarrow, v, B]}{\partial N_\downarrow}\right|_-\right) . \tag{57}$$

In Eqs.(56) and (57), it has been taken into account that increasing $N$ while $N_s$ is kept fixed is possible only if both $N_\uparrow$ and $N_\downarrow$ are increased, while increasing $N_s$ along a fixed $N$ can be done only if $N_\uparrow$ is increased and $N_\downarrow$ is decreased. Eq.(56) is in complete accordance with Eqs.(36a), (38a) and (39a). However, Eq.(57), with Eqs.(37a), (38a) and (39b), would give that the excitation energy from the $(M_\uparrow, M_\downarrow)$ lowest-energy state to the $(M_\uparrow + 1, M_\downarrow - 1)$ lowest-energy state equals $I_\downarrow - A_\uparrow$ (of the $(M_\uparrow, M_\downarrow)$ state), which is not true. That is, Eq.(57) is not valid at integer $N$'s (except for metals, with $I - A = 0$). This is because of the integer-$N$ derivative-discontinuity line between $E(M_\uparrow, M_\downarrow - 1)$ and $E(M_\uparrow + 1, M_\downarrow)$. Since there are no integer-$N_s$ derivative-discontinuity lines, Eq.(56) never breaks down. Transformations like Eqs.(56) and (57) between partial derivatives corresponding to different, rotated coordinate systems are valid fully only if the multivariable function in question is fully differentiable at the given point. It is worth mentioning that if there are no discontinuities in $E(N_\uparrow, N_\downarrow)$'s derivatives at some point, then there can be no discontinuities in $E(N, N_s)$'s derivatives either, since if there is no discontinuity in $\partial E(N_\uparrow, N_\downarrow)/\partial N_\sigma$ ($\sigma = \uparrow, \downarrow$), there is no discontinuity in $\partial E(N, N_s)/\partial N$, according to Eq.(56), but this means that Eq.(57) is valid, which then implies that there is no discontinuity in $\partial E(N, N_s)/\partial N_s$ (since there are no discontinuities in $\partial E(N_\uparrow, N_\downarrow)/\partial N_\sigma$) – this would give $v_{KS}^{s,+(-)}(\infty) = 0$ (see Eq.(70) below).

In the transformation from the $(N, N_s)$ representation to the $(N_\uparrow, N_\downarrow)$ representation, the connection between the energy derivatives emerges as

$$\left.\frac{\partial E[N_\uparrow, N_\downarrow, v, B]}{\partial N_\uparrow}\right|_+ = \left.\frac{\partial E[N, N_s, v, B]}{\partial N}\right|_+ + \left.\frac{\partial E[N, N_s, v, B]}{\partial N_s}\right|_+ \tag{58}$$

and

$$\left.\frac{\partial E[N_\uparrow, N_\downarrow, v, B]}{\partial N_\downarrow}\right|_+ = \left.\frac{\partial E[N, N_s, v, B]}{\partial N}\right|_+ - \left.\frac{\partial E[N, N_s, v, B]}{\partial N_s}\right|_- . \tag{59}$$

However, Eq.(58), with Eqs.(38a), (36a), and (37a), would give that the excitation energy from the $(M, M_s)$ lowest-energy state to the $(M, M_s + 2)$ lowest-energy state equals $A_\downarrow - A_\uparrow$, which is not true. This is because Eq.(58) does not hold for integer $N_\downarrow$'s due to the integer-$N_\downarrow$ derivative-discontinuity lines. In the case there is no integer-$N_\downarrow$ derivative-discontinuity line, $E(M, M_s + 2) - E(M, M_s) = E(M, M_s) - E(M+1, M_s - 1) - (E(M, M_s) - E(M+1, M_s + 1))$ becomes true. (For example, the energy surface for the carbon atom above the $(N, N_s)$-plane segment determined by (6,-2), (6,2), (7,3), and (7,-3) is constant; consequently, all energy differences above this segment will be zero, which makes the above equality trivially valid.) Similar considerations are valid in the case of Eq.(59), and for left-side $N_\sigma$-derivatives too.

The discontinuities of the derivatives of the energy lead to discontinuities in the KS potential. These are usually characterized by the differences in the potential's asymptotic value on the two sides of the discontinuities. Since the external and the classical Coulomb part of the KS potential have no discontinuity, the discontinuity of the KS potential is the discontinuity of the exchange-correlation potential. The discontinuity of the KS potential has proved to have physical significance, most notably in explaining the band gap problem of DFT [42] and in accounting for the correct dissociation of molecules [13,43]. Recently, increased attention has been focused on its investigation, both in ground-state and in time-dependent density functional theory; see [44,45], for example.

The discontinuities of the KS potential can be obtained on the basis of Eqs.(45) and (51). In the case of the spin-up and spin-down components, $v_{KS}^\sigma(\vec{r})$, this can be done similarly as in the spin-independent case. (We will apply a more elementary argument than the usual one, involving $\left.\frac{\delta T_s[n_\uparrow,n_\downarrow]}{\delta n_\sigma(\vec{r})}\right|_{+/-}$.) Write Eqs.(43) as

$$\varepsilon_{N_\sigma+1}^{\sigma,*} + v_{KS}^{\sigma,+}(\vec{r}^*) = -A_\sigma \qquad (60a)$$

and

$$\varepsilon_{N_\sigma}^{\sigma,*} + v_{KS}^{\sigma,-}(\vec{r}^*) = -I_\sigma \;, \qquad (60b)$$

where $\varepsilon_i^{\sigma,*}$ denote the KS energies corresponding to $v_{KS}^{\sigma,*}(\vec{r})$ for which $v_{KS}^{\sigma,*}(\vec{r}^*) = 0$ (the usual choice for $\vec{r}^*$ is $\infty$, assuming that the KS potential at infinity is the same in all directions). Then subtract Eq.(60b) from Eq.(60a) to find

$$\Delta_{xc}^\sigma \left(= v_{KS}^{\sigma,+}(\vec{r}) - v_{KS}^{\sigma,-}(\vec{r})\right) = (I_\sigma - A_\sigma) - \left(\varepsilon_{N_\sigma+1}^\sigma - \varepsilon_{N_\sigma}^\sigma\right) \qquad (61)$$

(where the * are suppressed; $\varepsilon_{N_\sigma}^\sigma$ and $\varepsilon_{N_\sigma+1}^\sigma$ are measured from a common point). Eq.(61) is the spin-resolved version of $\Delta_{xc} = (I-A) - (\varepsilon_{N+1} - \varepsilon_N)$, which explains the band gap problem of DFT (i.e., the difference between $I-A$ and the corresponding KS gap, $\varepsilon_{N+1} - \varepsilon_N$) through the discontinuity of the xc potential [42]. The difference between the left- and the right-side derivative of $T_s[n_\uparrow,n_\downarrow]$ with respect to $n_\sigma$ can be obtained simply on the basis of Eq.(41) (with Eqs.(8), (38), (39) and (61)), as

$$\left.\frac{\delta T_s[n_\uparrow,n_\downarrow]}{\delta n_\sigma(\vec{r})}\right|_+ - \left.\frac{\delta T_s[n_\uparrow,n_\downarrow]}{\delta n_\sigma(\vec{r})}\right|_- = \varepsilon_{N_\sigma+1}^\sigma - \varepsilon_{N_\sigma}^\sigma \;. \qquad (62)$$

We note here that to establish Eq.(61), or its spin-independent version, it is crucial that the left- and right-side KS potentials (or, the left- and the right-side derivative(s) of the single-particle kinetic-energy density functional) differ only by a constant – for which however only plausible arguments had been given [33a] until very recently. It has been shown in [19] (see also [46]) that the constant difference follows from the nature of the one-sided derivatives itself. Since the left- and the right-side derivative of a continuous functional $A[n]$ at a given point can be considered as the derivatives of two different fully differentiable functionals that are equal over the domain of $n(\vec{r})$ s with the given $N$, they can differ only by a constant (with respect to $\vec{r}$) [46]. (The explicit proof is given in the Appendix of [19].) All this, of course, is valid with the possible exceptions of zero-measure sets of the $\vec{r}$ space.

The discontinuities of the two components of the KS potential in the $(N, N_s)$ representation can be connected with the discontinuities of the two $v_{KS}^\sigma(\vec{r})$'s. Since Eq.(56), and more generally,

$$\left.\frac{\delta}{\delta n(\vec{r})}\right|_{+(-)} = \frac{1}{2}\left(\left.\frac{\delta}{\delta n_\uparrow(\vec{r})}\right|_{+(-)} + \left.\frac{\delta}{\delta n_\downarrow(\vec{r})}\right|_{+(-)}\right), \tag{63}$$

holds, $v_{KS}(\vec{r})$ is related with $v_{KS}^\sigma(\vec{r})$'s by

$$v_{KS}^\pm(\vec{r}) = \frac{1}{2}\left(v_{KS}^{\uparrow,\pm}(\vec{r}) + v_{KS}^{\downarrow,\pm}(\vec{r})\right). \tag{64}$$

Therefore, for the discontinuity $\Delta_{xc}\left(= v_{KS}^+(\vec{r}) - v_{KS}^-(\vec{r})\right)$ of $v_{KS}(\vec{r})$,

$$\Delta_{xc} = \frac{1}{2}\left(\Delta_{xc}^\uparrow + \Delta_{xc}^\downarrow\right) \tag{65}$$

emerges. With the use of Eq.(54), the left side of the discontinuity is found to be fixed as

$$v_{KS}^-(\infty) = 0 . \tag{66}$$

(This is the reason why $v_{KS}^-(\vec{r})$ was chosen to be inserted into Eqs.(48) to define the KS energies $\varepsilon_i^{\sigma,s+}$ and $\varepsilon_i^{\sigma,s-}$.)

In the case of the spin component of the KS potential, the above procedure does not work, since Eq.(57) does not hold; i.e., for density-functional derivatives,

$$\left.\frac{\delta}{\delta n_s(\vec{r})}\right|_{+(-)} \neq \frac{1}{2}\left(\left.\frac{\delta}{\delta n_\uparrow(\vec{r})}\right|_{+(-)} - \left.\frac{\delta}{\delta n_\downarrow(\vec{r})}\right|_{-(+)}\right) \tag{67}$$

(at integer $N$). Instead, by comparing the two sets of KS equations Eqs.(42) and (48), one obtains

$$\left(v_{KS}(\vec{r})+v_{KS}^s(\vec{r})\right)-v_{KS}^\uparrow(\vec{r})=\varepsilon_i^{\uparrow,s}-\varepsilon_i^\uparrow \tag{68a}$$

and

$$\left(v_{KS}(\vec{r})-v_{KS}^s(\vec{r})\right)-v_{KS}^\downarrow(\vec{r})=\varepsilon_j^{\downarrow,s}-\varepsilon_j^\downarrow \tag{68b}$$

(where $v_{KS}^s(\vec{r})$, $\varepsilon_i^{\uparrow,s}$ and $\varepsilon_j^{\downarrow,s}$ are +, or –, simultaneously, and the same holds for $v_{KS}^\uparrow(\vec{r})$ and $\varepsilon_i^\uparrow$, and for $v_{KS}^\downarrow(\vec{r})$ and $\varepsilon_j^\downarrow$). From Eq.(68), then,

$$v_{KS}^s(\vec{r})=\frac{1}{2}\left(\varepsilon_{N_\uparrow+1}^{\uparrow,s}-\varepsilon_{N_\downarrow}^{\downarrow,s}+v_{KS}^\uparrow(\vec{r})-\varepsilon_{N_\uparrow+1}^\uparrow-v_{KS}^\downarrow(\vec{r})+\varepsilon_{N_\downarrow}^\downarrow\right). \tag{69}$$

This, with the use of Eq.(51), and utilizing Eq.(54), yields

$$v_{KS}^{s,+}(\infty)=\frac{1}{2}\left(E(M,M_s+2)-E(M,M_s)-(\varepsilon_{M_\uparrow+1}^{\uparrow,-}-\varepsilon_{M_\downarrow}^{\downarrow,-})\right)$$

$$=\frac{1}{2}\left(E(M,M_s+2)-E(M,M_s)+A_\uparrow+v_{KS}^{\uparrow,+}(\infty)-v_{KS}^{\uparrow,-}(\infty)-I_\downarrow\right) \tag{70a}$$

and

$$v_{KS}^{s,-}(\infty)=-\frac{1}{2}\left(E(M,M_s-2)-E(M,M_s)-(\varepsilon_{M_\downarrow+1}^{\downarrow,-}-\varepsilon_{M_\uparrow}^{\uparrow,-})\right)$$

$$=-\frac{1}{2}\left(E(M,M_s-2)-E(M,M_s)+A_\downarrow+v_{KS}^{\downarrow,+}(\infty)-v_{KS}^{\downarrow,-}(\infty)-I_\uparrow\right) \tag{70b}$$

(note that $v_{KS}^{\uparrow(\downarrow),-}(\infty)=0$). To obtain the second equality in Eqs.(70), Eq.(45) has been utilized. It can be seen that $v_{KS}^{s,+(-)}(\infty)-\frac{1}{2}\left(v_{KS}^{\uparrow,+(-)}(\infty)-v_{KS}^{\downarrow,-(+)}(\infty)\right)\neq 0$, in accordance with the energy derivative discontinuity analyzed below Eq.(57). Subtracting Eq.(70b) from Eq.(70a) gives

$$\Delta_{xc}^s=\frac{1}{2}\left(E(M,M_s+2)-E(M,M_s)+E(M,M_s-2)-E(M,M_s)-(\varepsilon_{M_\uparrow+1}^{\uparrow,-}-\varepsilon_{M_\downarrow}^{\downarrow,-}+\varepsilon_{M_\downarrow+1}^{\downarrow,-}-\varepsilon_{M_\uparrow}^{\uparrow,-})\right)$$

$$=\frac{1}{2}\left(E(M,M_s+2)-E(M,M_s)+E(M,M_s-2)-E(M,M_s)+A_\uparrow-I_\uparrow+A_\downarrow-I_\downarrow+\Delta_{xc}^\uparrow+\Delta_{xc}^\downarrow\right). \tag{71}$$

Comparing this result with $\Delta_{xc}^s=\frac{1}{2}\left(\Delta_{xc}^\uparrow+\Delta_{xc}^\downarrow\right)$, which would emerge if Eq.(57) (and the corresponding equation with the signs in the subscripts changed to the opposite) held, shows the effect of the derivative discontinuity at integer *N*'s (because of which Eq.(57) does not hold).

Eq.(70) can be obtained also by following similar arguments as in the case of Eq.(61) previously: Write the KS energies in Eqs.(51) as $\varepsilon_{M_\uparrow+1}^{\uparrow,s+}=\varepsilon_{M_\uparrow+1}^{\uparrow,*}+v_{KS}^{s,+}(\vec{r}^*)$, $\varepsilon_{M_\downarrow}^{\downarrow,s+}=\varepsilon_{M_\downarrow}^{\downarrow,*}-v_{KS}^{s,+}(\vec{r}^*)$, $\varepsilon_{M_\uparrow}^{\uparrow,s-}=\varepsilon_{M_\uparrow}^{\uparrow,*}+v_{KS}^{s,-}(\vec{r}^*)$, and $\varepsilon_{M_\downarrow+1}^{\downarrow,s-}=\varepsilon_{M_\downarrow+1}^{\downarrow,*}-v_{KS}^{s,-}(\vec{r}^*)$, then subtract the two equations from each other. The discontinuity in $T_s[n,n_s]$'s derivative with respect to $n_s(\vec{r})$ emerges, on the basis of Eq.(47b), using Eqs.(7b), (37), and (71), as

$$\left.\frac{\delta T_s[n,n_s]}{\delta n_s(\vec{r})}\right|_+ - \left.\frac{\delta T_s[n,n_s]}{\delta n_s(\vec{r})}\right|_- = \frac{1}{2}\left(\varepsilon_{M_\uparrow+1}^{\uparrow,-} - \varepsilon_{M_\downarrow}^{\downarrow,-} + \varepsilon_{M_\downarrow+1}^{\downarrow,-} - \varepsilon_{M_\uparrow}^{\uparrow,-}\right)$$

$$= \frac{1}{2}\left(\left.\frac{\delta T_s[n_\uparrow,n_\downarrow]}{\delta n_\uparrow(\vec{r})}\right|_+ - \left.\frac{\delta T_s[n_\uparrow,n_\downarrow]}{\delta n_\uparrow(\vec{r})}\right|_- + \left.\frac{\delta T_s[n_\uparrow,n_\downarrow]}{\delta n_\downarrow(\vec{r})}\right|_+ - \left.\frac{\delta T_s[n_\uparrow,n_\downarrow]}{\delta n_\downarrow(\vec{r})}\right|_-\right). \quad (72)$$

Notice that Eq.(72) has emerged in spite of Eq.(67). This has the consequence that

$$\left.\frac{\delta T_s[n,n_s]}{\delta n(\vec{r})}\right|_+ - \left.\frac{\delta T_s[n,n_s]}{\delta n(\vec{r})}\right|_- = \left.\frac{\delta T_s[n,n_s]}{\delta n_s(\vec{r})}\right|_+ - \left.\frac{\delta T_s[n,n_s]}{\delta n_s(\vec{r})}\right|_-, \quad (73)$$

since Eq.(63) gives $\left.\frac{\delta T_s[n,n_s]}{\delta n(\vec{r})}\right|_+ - \left.\frac{\delta T_s[n,n_s]}{\delta n(\vec{r})}\right|_-$ just as the expression in the second line of Eq.(72).

Finally, we mention that since the submission of the present paper, a related work by Capelle et al. has appeared [47]. Capelle et al. investigate the differences between the spin flip energy $E(M,M_s+2)-E(M,M_s)$ [$E(M,M_s-2)-E(M,M_s)$] and the corresponding KS value $\varepsilon_{M_\uparrow+1}^\uparrow - \varepsilon_{M_\downarrow}^\downarrow$ [$\varepsilon_{M_\downarrow+1}^\downarrow - \varepsilon_{M_\uparrow}^\uparrow$], and between the "fundamental spin gap" $E(M,M_s+2)-E(M,M_s)+E(M,M_s-2)-E(M,M_s)$ and the corresponding KS gap. They explain the differences between the real and the KS spin flip energies with the help of the ensemble theory of DFT for excited states [48], attributing them to a kind of derivative discontinuity [49] of the excited-state-ensemble xc functional $E_{xc}^w[n_\uparrow^w,n_\downarrow^w]$ at $w=0$. They also determine these differences in terms of corrections to $\varepsilon_{M_\uparrow+1}^\uparrow - \varepsilon_{M_\downarrow}^\downarrow$, or $\varepsilon_{M_\downarrow+1}^\downarrow - \varepsilon_{M_\uparrow}^\uparrow$, from the single-pole approximation of time-dependent DFT for the calculation of excitation energies [50]. In the present work, we have shown how the differences in question (denoted by $\Delta_{xc}^{sf+}$ and $\Delta_{xc}^{sf-}$ in [47]) can be explained within the framework of spin-polarized DFT, in the spirit of Perdew et al. [13]. We have found that they can be attributed to a nonzero asymptotic value of the spin component of the SDFT KS potential (namely, $2v_{KS}^{s,+}(\infty)$ and $-2v_{KS}^{s,-}(\infty)$; see Eq.(70)), instead of being xc derivative discontinuities. It is then the difference between the real and the KS fundamental spin gap, $\Delta_{xc}^{sf+}+\Delta_{xc}^{sf-}$, what emerges as (double of) a derivative discontinuity, namely, as $2\Delta_{xc}^s[=2(v_{KS}^{s,+}(\infty)-v_{KS}^{s,-}(\infty))]$ (note that in [47], $\Delta_{xc}^s$ denotes $\Delta_{xc}^{sf+}+\Delta_{xc}^{sf-}$).

## IV. Summary

Formulae describing the energy surface $E(N, N_s)$ of spin-polarized density functional theory, defined by Eq.(12), have been presented, and the shape of the surface has been displayed. Based on these results, the negatives of the left/right-side derivatives of the energy with respect to $N$, $N_\uparrow$, and $N_\downarrow$ have been shown to give the fixed-$N_s$, spin-up, and spin-down ionization potentials/electron affinities, respectively, while the derivative of $E[N, N_s, v, B]$ with respect to $N_s$ gives the (signed) half excitation energy to a state with $N_s$ increased, or decreased, by 2 – provided the convexity of $E_{l.s.}(M, M_s)$ with respect to $M$ and $M_s$, with $E_{l.s.}(M, M_s)$ denoting the lowest-lying energy-eigenstate with integer particle number $M$ and integer spin number $M_s$ that is also a particle-number and a spin eigenstate. (The cases where this convexity condition does not hold have also been discussed.) The highest occupied and lowest unoccupied Kohn-Sham spin-orbital energies have been identified as the corresponding spin-up and spin-down ionization potentials and electron affinities, if calculated with the use of the left-side, or the right-side, Kohn-Sham spin-potentials, respectively. On the other hand, the excitation energies to the states with $M_s \pm 2$ can be obtained as the differences between the lowest unoccupied and the opposite-spin highest occupied spin-orbital energies, if the $(N, N_s)$ representation of the Kohn-Sham spin-potentials is used. Besides leading to real excitation energies, the use of the latter spin-potentials has another advantageous consequence: It restores the Aufbau principle in spin-polarized Kohn-Sham calculations. The discontinuities of the energy derivatives and the Kohn-Sham potentials, which are essential to account for in the development of accurate energy density functionals, have been analyzed and related.

**Acknowledgments:** Useful discussion with Jan Moens is acknowledged. T.G. acknowledges a visiting fellowship from the Fund for Scientific Research – Flanders (FWO).

## Appendix A: Convexity of $E(M, M_s)$ for noninteracting systems

The energy of a noninteracting-electron system in a given $\left(v^\uparrow(\vec{r}), v^\downarrow(\vec{r})\right)$ is

$$E(M_\uparrow, M_\downarrow) = \sum_{i=1}^{M_\uparrow} \varepsilon_i^\uparrow + \sum_{i=1}^{M_\downarrow} \varepsilon_i^\downarrow \ . \tag{A1}$$

Since $\varepsilon_i^\sigma$ follow each other in a monotonously increasing order for both spin indices,

$$2E(M_\uparrow, M_\downarrow) \leq E(M_\uparrow - 1, M_\downarrow) + E(M_\uparrow + 1, M_\downarrow) \tag{A2a}$$

and

$$2E(M_\uparrow, M_\downarrow) \leq E(M_\uparrow, M_\downarrow - 1) + E(M_\uparrow, M_\downarrow + 1) \ , \tag{A2b}$$

which means that $E(M_\uparrow, M_\downarrow)$ is convex both in $M_\uparrow$ and in $M_\downarrow$. With the use of $E(M \pm 2, M_s) = E(M_\uparrow \pm 1, M_\downarrow \pm 1) = E(M_\uparrow \pm 1, M_\downarrow) + E(M_\uparrow, M_\downarrow \pm 1) - E(M_\uparrow, M_\downarrow)$ (which follows from Eq.(A1)), the sum of Eq.(A2a) and Eq.(A2b) yields

$$2E(M, M_s) \leq E(M - 2, M_s) + E(M + 2, M_s) \ . \tag{A3a}$$

Since $E(M, M_s \pm 2) = E(M_\uparrow \pm 1, M_\downarrow \mp 1) = E(M_\uparrow \pm 1, M_\downarrow) + E(M_\uparrow, M_\downarrow \mp 1) - E(M_\uparrow, M_\downarrow)$ holds, too, the sum of Eq.(A2a) and Eq.(A2b), but now with a different grouping of the terms, gives also

$$2E(M, M_s) \leq E(M, M_s - 2) + E(M, M_s + 2) \ . \tag{A3b}$$

Eqs.(A3) then mean that $E(M, M_s)$ is convex in $M$, and in $M_s$. (At fixed spin number/particle number, the particle number/spin number can change only by two, if whole particles are allowed only.) Note that the proof itself allows holes below the highest-occupied spin-orbitals, but their filling is not allowed as going from $N$ particles to $N+1$ particles. However, since the energy is actually defined by the lowest-energy state with a given $N$ and $N_s$, holes are excluded.

## Appendix B: Equality of interacting and Kohn-Sham chemical potentials

The equality of the chemical potential of an interacting electron system with that of the corresponding KS noninteracting system is an inherent part of the beauty of the KS idea. As often noted in the DFT literature, the equality emerges simply by construction (see e.g. [24]). Since it plays a key role in obtaining Eqs.(43), (44), (49) and (50), in this Appendix we provide a simple argument to show its validity. For simplicity, we consider the spin-independent case.

If a $n(\vec{r})$ is the ground-state density in some $v(\vec{r})$ in the interacting case, it satisfies

$$\frac{\delta F[n]}{\delta n(\vec{r})} + v(\vec{r}) = \mu[N,v] \; , \tag{B1}$$

where

$$\mu[N,v] = \frac{\partial E[N,v]}{\partial N} \; . \tag{B2}$$

Similarly, if a $\rho(\vec{r})$ is the ground-state density in some $w(\vec{r})$ in the noninteracting case, it satisfies

$$\frac{\delta T_s[\rho]}{\delta \rho(\vec{r})} + w(\vec{r}) = \mu_{s.p.}[N,w] \; , \tag{B3}$$

where

$$\mu_{s.p.}[N,w] = \frac{\partial E^{s.p.}[N,w]}{\partial N} \; . \tag{B4}$$

Both of the above Euler equations can be established with allowing potentials with arbitrary asymptotic values, extending the usual domain of DFT. Then $v[n]$, and $w[\rho]$, will give a class of potentials (instead of one potential), which differ only by an overall constant. But $v(\vec{r}) - \mu$, and $w(\vec{r}) - \mu_{s.p.}$, will be unique; note that $\mu[N, v+c] = \mu[N,v] + c$. (Of course, if there is a discontinuity in the energy derivative with respect to the particle number, there will be a left-side and a right-side version of the equations, and the chemical potentials.)

Now, let $n(\vec{r})$ be a ground-state density corresponding to some $v(\vec{r})$ in the interacting case. Eq.(B1) can be rewritten as

$$\frac{\delta T_s[n]}{\delta n(\vec{r})} + v_{KS}(\vec{r}) = \mu \; , \tag{B5}$$

with

$$v_{KS}(\vec{r}) \doteq \frac{\delta(F[n] - T_s[n])}{\delta n(\vec{r})} + v(\vec{r}) \; . \tag{B6}$$

If $n(\vec{r})$ is noninteracting-w-representable, Eq.(B5) simply means that $n(\vec{r})$ corresponds to $w(\vec{r}) = v_{KS}(\vec{r})$ in the noninteracting case, with $\mu_{s.p.}[N, v_{KS}] = \mu[N,v]$. That is,

$$\frac{\partial E[N,v]}{\partial N} = \frac{\partial E_{s.p.}[N, v_{KS}]}{\partial N} \; . \tag{B7}$$

The spin-resolved version is analogous.

We note that if one wishes to remain in the usual DFT domain of external potentials with zero asymptotic constant, Eq.(B7) can still be obtained – though in a less elegant way, without establishing an equality of the chemical potentials. Separating the asymptotic constant in $v_{KS}(\vec{r})$, $v_{KS}(\vec{r}) = v_{KS}^0(\vec{r}) + v_{KS}(\infty)$ (assuming the constancy of $v_{KS}(\infty)$), Eq.(B5) can be rewritten as

$$\frac{\delta T_s[n]}{\delta n(\vec{r})} + v_{KS}^0(\vec{r}) = \mu[N,v] - v_{KS}(\infty) \ . \tag{B8}$$

Then $v_{KS}^0(\vec{r}) = w(\vec{r})$, and $\mu[N,v] - v_{KS}(\infty) = \mu_{s.p.}[N, v_{KS}^0]$. From the latter,

$$\mu = \frac{\partial E_{s.p.}[N, v_{KS}^0]}{\partial N} + v_{KS}(\infty) = \frac{\partial E_{s.p.}[N, v_{KS}^0 + v_{KS}(\infty)]}{\partial N} = \frac{\partial E_{s.p.}[N, v_{KS}]}{\partial N} \tag{B9}$$

emerges, giving back Eq.(B7) in the end.